\documentclass{aa}
\usepackage{graphicx}
\usepackage{txfonts}
\usepackage{natbib}
\usepackage[english]{babel}
\usepackage{placeins}

\begin{document}

\title{Observations of UX Ori in deep minima with the Nordic Optical Telescope. I. Analysis of spectral lines}
\titlerunning{Observations of UX Ori in deep minima. Analysis of spectra}
\authorrunning{Tambovtseva et al.}

\author{
  L.V. Tambovtseva\inst{1}, A.A. Djupvik\inst{2,3}, V.P. Grinin\inst{1},
  H. Weber\inst{4}, H. Bengtsson\inst{5}, H. De Angelis\inst{5},
  G. Duszanowicz\inst{5}, D. Heinonen\inst{5}, L. Hermansson\inst{5},
  G. Holmberg\inst{5},
  T. Karlsson\inst{5}, M. Larsson\inst{5}, J. Warell\inst{5}, T. Wikander\inst{5}
}

\institute {
  Pulkovo Astronomical Observatory, Russian Academy of Sciences,
  Pulkovskoe sh. 65, St. Petersburg, 196140, Russia \\
                 \email{lvtamb@mail.ru}
                 \and
                 Nordic Optical Telescope, Rambla Jos\'{e} Ana Fern\'{a}ndez P\'{e}rez 7, ES-38711 Bre\~{n}a Baja, Spain
                 \and
                 Department of Physics and Astronomy, Aarhus University, Munkegade 120, DK-8000 Aarhus C, Denmark
                 \and
                 Lule\aa \ University of Technology, SE-971 87 Lule\aa \, Sweden
                 \and
                 SAAF, Svensk Amat\"{o}rAstronomisk F\"{o}rening, Sweden
}

\date{Received; accepted }

\abstract
    {UX Orionis stars are the most active young stars; they undergo
        sporadic fadings of 2 - 4 magnitudes in the V-band, due
to variable circumstellar extinction caused by a nearly edge-on
star--disc system. The long-lasting monitoring of a number of
stars of this type with the Nordic Optical Telescope from 2019 to
2024 has given a rich collection of material of high-resolution (R
$\sim$ 25000) spectra obtained during different brightness states
of the stars. In this paper, we present the results of
observations for UX Ori itself. Until now only one spectrum of
high resolution had been obtained for this star during brightness
minimum, making it difficult to do a comprehensive analysis.}
{Our aim is to analyse how different spectral lines change during
such irregular fading events, when the star is going in and out of
eclipses, obscured by dust along the line of sight.}
{For this purpose we provide a comparative analysis of the
profiles and equivalent widths of the spectral lines belonging to
the different atoms and ions. In addition we compare the results
for UX~Ori with those made for another target in our sample:
RR~Tau.}
{Common features of variability are revealed: (1) a strengthening
of the H$\alpha$ line relatively to the continuum during eclipses;
(2) the appearance of additional emission on the frequencies of
photospheric lines (e.g. FeII, CaII, SiII). The different
behaviour of the spectral lines during fading found for UX Ori and
RR Tau may be caused by two effects: a different contribution of
the scattered light to the stellar flux during eclipses or a less
intense disc wind of UX Ori.}
{}


\maketitle

\section{Introduction}

UX Orionis stars (UXORs) are among the most active young stars.
This is demonstrated by irregular flux variability, which takes
the form of sporadic brightness weakening with amplitudes from 2
to 4 magnitudes in the V-band. The UXOR family includes both
low-mass T Tauri stars (TTSs) and intermediate-mass Herbig Ae
stars (HAEs). The latter makes up the majority of this family. The
typical duration of these brightness minima ranges from a few days
to a few weeks for stars of spectral type  A - F \citep[see
e.g.][and references therein]{melnikov2001,shakhovskoi2003}.
Cooler TTSs often exhibit small-amplitude brightness minima
(dippers) with duration of 1 - 2 days \citep[see e.g.][and
references therein]{ansdell2016,roggero2021}. Occasionally, very
long-lasting minima (more than a year) are observed \citep[see
e.g.][]{rostopchina2012}. This photometric behaviour is due to
variable circumstellar (CS) extinction caused by dust from their
protoplanetary discs, which in these stars make a small angle with
the line of sight (LOS). In other words, an observer sees these
star--disc systems nearly edge-on. This conclusion has been
reached based (1) on observations of linear polarisation of UXORs
\citep{gri91} and their modelling \citep{natta2000,shulman2019},
(2) on the behaviour of the shape of the H$\alpha$ line profiles
with the level of photometric activity \citep{gri1996,vioque18},
and (3) on the modelling of hydrogen lines \citep{tam20}. Modern
interferometric observations have confirmed this conclusion
\citep[and refereces therein]{kre13,kreplin2016}.

Nevertheless, while the physical explanation of the flux fading is found
  to be dust fragments passing in front of the star and eclipsing it, the cause
  of these sporadic appearances of dust clouds is less well established.
  One suggestion is that the disc wind or the poloidal component of the magnetic
  field is lifting up material from the disc atmosphere.  In order to
  investigate how disc winds, dust clouds, and sub-structure in the inner regions of
  protoplanetary discs are connected, we have monitored with high-resolution
  spectroscopy several UXOR stars. Dramatic changes in the emission
  lines during the flux variation was shown by \citet{rod02} for the
  UXOR star RR~Tau, and hybrid models treating both accretion and magneto-centrifugal
  disc winds have been proposed to explain them \citep{tam20}. Thus, monitoring the
  spectral behaviour of such stars during the dust eclipses is a way to investigate the
  processes at play. In addition, since the dust eclipses behave as a coronagraph, these
  observations offer a view into the inner regions of young protoplanetary systems that
  cannot be resolved spatially.

Echelle spectra of high resolution (R $\sim$ 25000 \AA) were
obtained with the Nordic Optical Telescope (NOT) of several UXORs
over the period 2019 - 2024, covering different brightness states
of several UXORs, including deep minima. These observations were
possible thanks to the flexible Target-of-Opportunity mode offered
by the NOT and to knowing when to trigger based on photometric
monitoring from our network of amateur astronomers. The first
results, the spectra of the UXOR star RR~Tau, were analysed and
presented together with the modelling of its dust eclipses in
\citet{grinin2023}.

In the present paper (Part I) we give a detailed analysis of the
behaviour of the main spectral lines of UX~Ori, the star that
gives its name to this whole group of young stars. It should be
  mentioned that until now only one spectrum in the H$\alpha$ region
has been observed at high spectral resolution when UX Ori was in
the brightness minimum \citep{gri94}. All other spectral
observations of this star were made in its bright state.
Therefore, the spectra obtained in the present project with the
NOT are unique because they were obtained in different states of
brightness including the weakest ones. This permitted us for the
first time to compare the behaviour of the lines of different
atoms at the different ionisation stages. For such an analysis we
selected the lines that belong to both neutral and ionised atoms.
In addition, we selected the most interesting moments when the
lines demonstrate dramatic or unexpected profile changes and
considered them separately.

In addition to the study of the main spectral lines we aimed to
investigate the reaction of the diffuse interstellar band (DIB)
6283 \AA \ during brightness minima. \citet{rod02} reported
changes in the DIB in RR~Tau spectra during eclipses. Thus, a
suspicion arose that the DIB can absorb the radiation of the star
in the CS envelope. However, the spectra were not of high quality.
Therefore, it was important to understand whether the DIB is
strengthened during eclipses, and for this purpose UXORs are
suitable because they are expected to display strong changes with
the eclipse amplitude.

We compared the results of these observations with those obtained
for RR~Tau. In Sect.~\ref{obs} we describe the observations and
the data reduction. In Sect.~\ref{bright-faint} we present the
spectra during the bright and faint phases, various special
features are described in Sect.~\ref{special}, and a comparison of
spectral behaviour during eclipses between UX~Ori and RR~Tau is
shown in Sect.~\ref{comparison}. In Sect.~\ref{discussion} we
discuss the implications  and possible interpretations of the
results. Modelling of the line profiles and eclipses for UX~Ori
will be presented in the next paper of the series (Part II).

  \begin{figure*}
  \center
  \includegraphics[width=17cm]{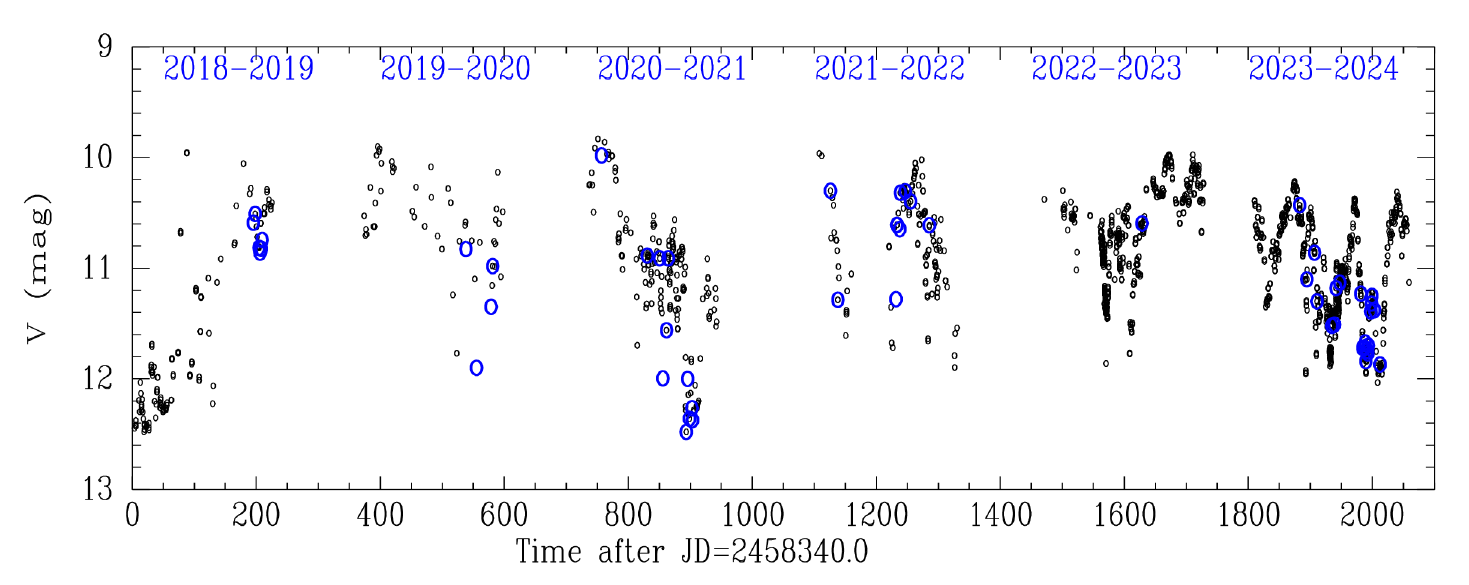}
  \caption{V-band observations of UX~Ori (black small circles) obtained
mainly from observers (e.g. SAAF) and with a few additional data
points from the NOT, are shown together with the epochs of FIES
spectroscopy (large blue circles). When quasi-simultaneous SAAF or
NOT V-band is lacking (empty blue circles), the nearest
photometric point found by searching the AAVSO database is shown
in Table~\ref{tab:fies-obs}.}
  \label{fig-saaf}
\end{figure*}

\section{Observations}
\label{obs}

The observations were obtained with the high-resolution FIber-fed
Echelle Spectrograph (FIES) through its low-resolution fibre
\citep{tel14} at the Nordic Optical Telescope \citep{dju10} over a
period from 2019 February to 2024 March. Most of the data was
obtained through observing programmes at the Nordic Optical
Telescope (NOT) that were awarded Target-of-Opportunity time, a
scheduling flexibility needed in order to catch the unpredictable
variability of UXOR stars during their irregular fading events.
Another necessary requirement was the frequent photometric
monitoring and rapid communication of alerts when a fading was
about to occur for one of our target stars. This was provided by
the Swedish Amateur Astronomy Association (SAAF).\footnote{Svensk
Amat\"{o}rAstronomisk F\"{o}rening.} The collaboration with SAAF
and the observing strategy is described more in detail in
\citet{grinin2023}.

FIES can be used with different fibres to obtain different
spectral resolutions up to R $\sim$ 67000, but we selected the
low-resolution fibre, giving a resolving power of R $\sim$ 25000
in order to obtain a reasonable S/N even when the targets went
into fading. For UX~Ori a nominal exposure time of 1800 seconds
was used initially and until re-coating of the spectrograph
mirrors took place in 2021 July, after which the typical exposure
time was lowered to 1200 s due to an increase in the spectrograph
throughput. The wavelength coverage is around 3700-9100 \AA. Data
reduction was made with the pipeline FIEStool\footnote{FIEStool
manual, Stempels 2005, Nordic Optical Telescope, see
\url{https://www.not.iac.es/instruments/fies/}}
\citep{stempels2017}. We used the tool {\em molecfit}
\citep{smette2015} in its version 1.5.9 to correct the spectra for
telluric contamination from the molecules H$_2$O, O$_3$, and
O$_2$. All the 1D spectra were normalised and put on a
heliocentric velocity grid for further measurements and analysis
using standard tasks in the IRAF package.

An overview of the observations is given in
Table~\ref{tab:fies-obs} together with the quasi-simultaneous
brightness obtained by photometry supplied from various sources:
1) The SAAF V-band, 2) NOT/ALFOSC or NOT/StanCam V-band, 3)
AAVSO\footnote{The American Association of Variable Star
Observers.} V-band, 4) ASASSN\footnote{All Sky Automated Survey
for Supernovae.} g-band, or 5) AAVSO Visual magnitudes, where the
latter two were used when no proper V-band magnitude was available
within 1 day of the FIES spectrum. Figure~\ref{fig-saaf} shows the
all collected V-band monitoring by SAAF, with the 54 epochs of
FIES spectra marked in blue. For spectra where quasi-simultaneous
SAAF or NOT V-band photometry was not obtained, we searched the
AAVSO database for the nearest (in time) photometric point in
either of the bands V, g, or Visual (see
Table~\ref{tab:fies-obs}).

\section{Spectra in the bright and faint states of the star}
\label{bright-faint}

The young star UX~Ori is a Herbig~Ae star of spectral type A3III.
We took into account the investigation by the Crimean group
accumulated in the paper by \citet{rostopchina1999}, who
determined the fundamental parameters of 11 UXOR stars
(luminosities, masses, radii and ages) on the basis of long-term
photometric monitoring, argueing that they are young stars, as
well as the corrected parameters of UX Ori from the latest GAIA
data \citep{guzman2021}. We adopt the following stellar
parameters: $M_*=2M_\odot$, R$_*=1.9R_\odot$, T$_{eff}=8500~$K,
and $\log g=4$. The rotational velocity $v \sin i$ = 140 km
s$^{-1}$, the radial velocity RV = + 18 km s$^{-1}$
\citep{gri2001}.

\begin{figure*}
\hspace{-0.4cm}\includegraphics[width=41mm]{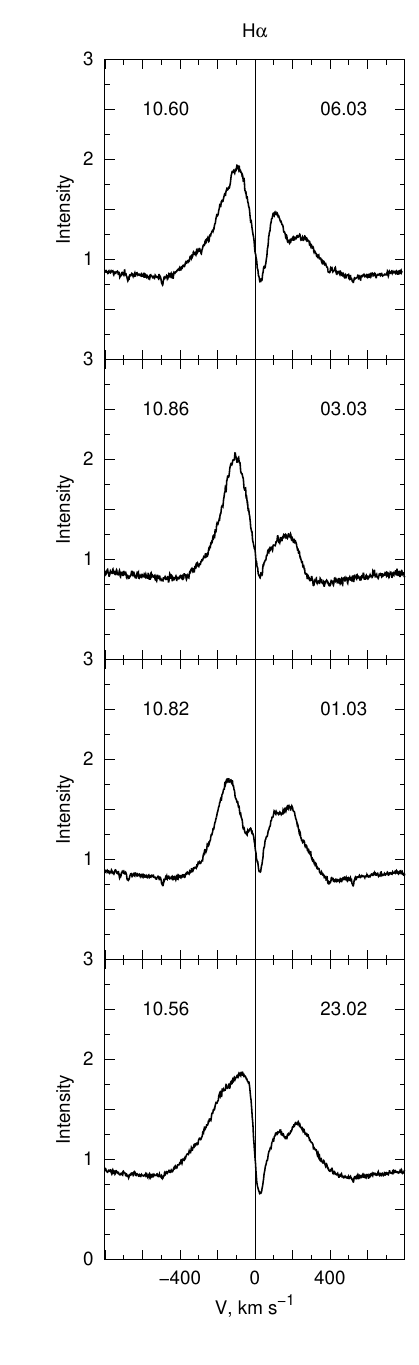}
\hspace{-0.6cm}\includegraphics[width=41mm]{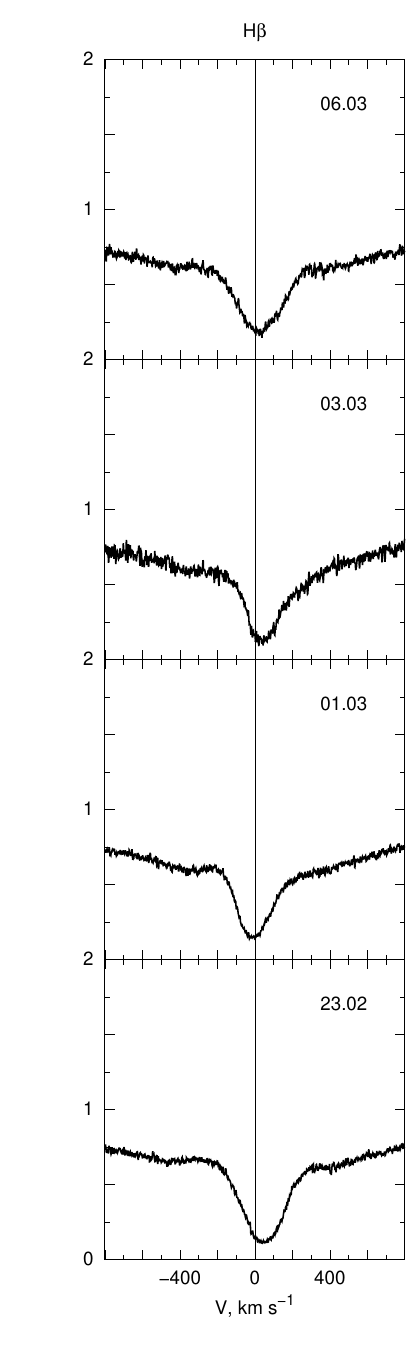}
\hspace{-0.6cm}\includegraphics[width=41mm]{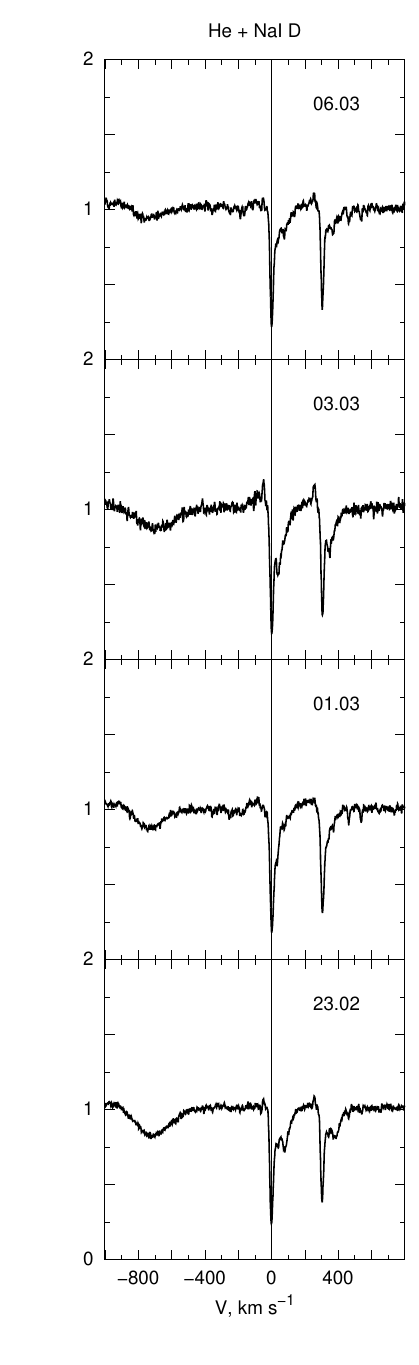}
\hspace{-0.6cm}\includegraphics[width=41mm]{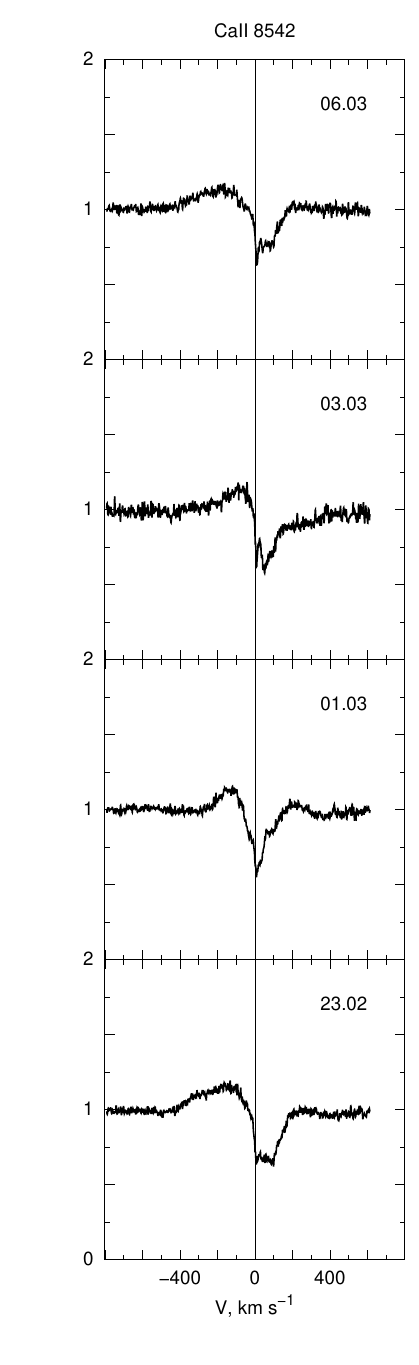}
\hspace{-0.6cm}\includegraphics[width=41mm]{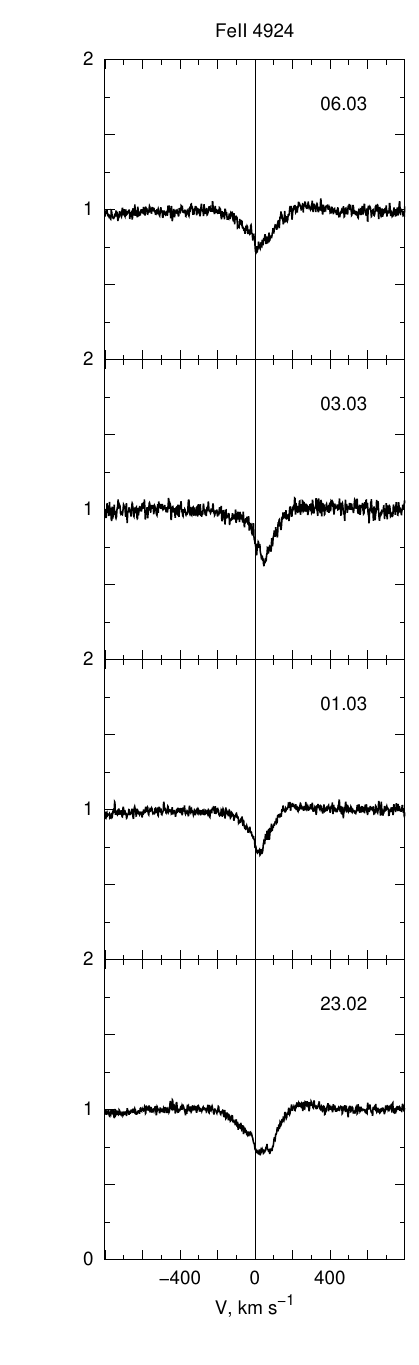}
\caption{UX~Ori line profiles of H$\alpha$, H$\beta$, NaI~D,
  CaII~8542~\AA, and FeII~4924 \AA \ during normal bright states
  in 2019. The line profiles are shown with respect to the stellar
  radial velocity (vertical line).
The dates of the observations are shown in each plot. The left
panel includes the values of the V-magnitudes, which are related
to all the lines. The narrow emission components in the sodium
lines are telluric lines (see text for details).} \label{fig2}
\end{figure*}

In this section we present and analyse the line profiles of the
hydrogen (H$\alpha$, H$\beta$), helium (HeI 5876 \AA), sodium (NaI
D 5889 \AA \ (D$_1$), 5895 \AA \ (D$_2$)), calcium (CaII 8542 \AA)
and iron (FeII 4925 \AA) lines at different brightness states of
the star over the long-lasting observational period from 2019 to
2024. When the star was in its normal (bright) state in 2019 is
shown in Fig.~\ref{fig2}. As the star was entering a short
brightness minimum in the beginning of December 2020 and
{\em(possibly)} going out of it is shown in Fig.~\ref{fig3}. The
brightness minimum in January 2021 is shown in Fig.~\ref{fig4}.
The last season of our observations (2023 - 2024) is presented in
the following two figures. Fig.~\ref{fig5} shows the beginning of
the minimum in November-December 2023, and Fig.~\ref{fig6}
presents the brightness minimum in January 2024. The line profiles
are given in the star coordinate system.

The most prominent hydrogen lines H$\alpha$ and H$\beta$ are the
most informative lines regarding conditions in the close vicinity
of the young star. In the normal (bright) state, the H$\alpha$
line has a double-peaked profile with a typical ratio V/R $> 1$
 \citep[][and these observations]{gri2001}. Earlier only once has a
single-peaked H$\alpha$ emission profile been obtained during the
deep and long-lasting brightness minima \citep{gri94}. Our current
observations demonstrate a single-peaked profile during the
eclipse in January 2021 (Fig.~\ref{fig4}) and the long-lasting (a
month or more) eclipse in January 2024 (Fig.~\ref{fig6}). These
observations suggest that the single-peaked and almost symmetric
H$\alpha$ profile is a typical shape during the deep brightness
minima.

In the normal state the H$\beta$ line profile shows a red-shifted
absorption component; this profile becomes wider and shallower
during the brightness minima, an indication of accretion of the
material (see e.g. Figs.~\ref{fig4} and \ref{fig6}).

The HeI 5876 \AA \ line does not disappear in the brightness
minima as observed in the case of the RR Tau eclipses. The
behaviour of the sodium lines also does not correlate with the
brightness.

In Figs.~\ref{fig2}, \ref{fig3}, \ref{fig4}, \ref{fig5} and
\ref{fig6} the calcium lines CaII~8542~\AA \ and iron
FeII~4924~\AA \ (one line of the multiplet) are also shown. These
lines are mainly photospheric, but they are interesting because
they also contain a circumstellar line which can be both in
absorption and in emission (so-called CS veiling). An analysis of
the modification of these profiles is helpful to understand the
physical processes in the nearest vicinity of the star. As is seen
from Fig.~\ref{fig2}, in the normal state, the lines of ionised
calcium are of the inverse P~Cygni type, which is typical for
accreting matter. In the faint
  state on the other hand Figs.~\ref{fig4}, \ref{fig5}, and
\ref{fig6} show that an additional emission appears and the
absorption may partially or completely disappear. Sometimes an
additional absorption appears on the red side (see
Fig.~\ref{fig3}, and \ref{fig4}). In the latter case it is not
connected with changes in brightness.\footnote{We observe the same
in NaI~D. In general, NaI~D and CaII do not correlate in
behaviour, except for a few days (Fig.~\ref{fig2})} The FeII line
behaves in a similar way as the CaII line: when the star is at
maximum brightness it is in absorption and red-shifted, and during
the minima an emission component is added, and the line becomes
wider and shallower (see e.g. Fig.~\ref{fig5} for the date Nov 30,
2023). We can suggest the following explanation for the changes
observed in these line profiles during eclipses. When the star
enters minimum, the emission begins to reveal itself on the
background of the decreasing continuum. This emission is always
present and it forms in a region which is not obscured by the
screen. In the case of RR~Tau we estimated the velocity of the
blue-shifted emission in the photospheric lines (e.g. FeII) and
suggested that it is due to the disc wind on the periphery of the
disc \citep{grinin2023}. The same may be suggested for UX Ori but
keeping in mind changes in the line profiles, the scale of the
lines variation is different and not as significant as in the case
of RR~Tau.

From the H$\alpha$ emission lines obtained in the bright
  state of the star one can estimate a gas velocity of about 400~km~s$^{-1}$;
  the helium and sodium lines indicate gas velocities $\sim$ 200~km~s$^{-1}$
  (see Fig.~\ref{fig2}.)

\section{Special features of the spectra}
\label{special}

\subsection{Deviation of the central dip in different epochs}

\begin{figure*}
\centering
\includegraphics[width=60mm]{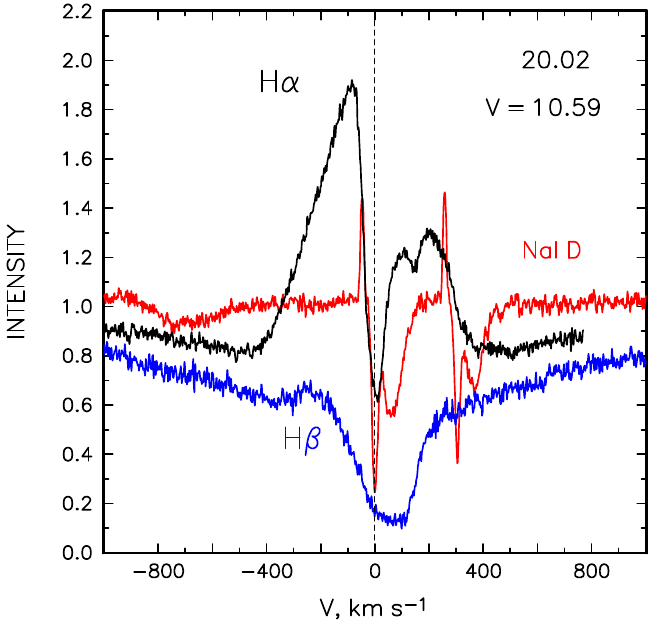}
\includegraphics[width=60mm]{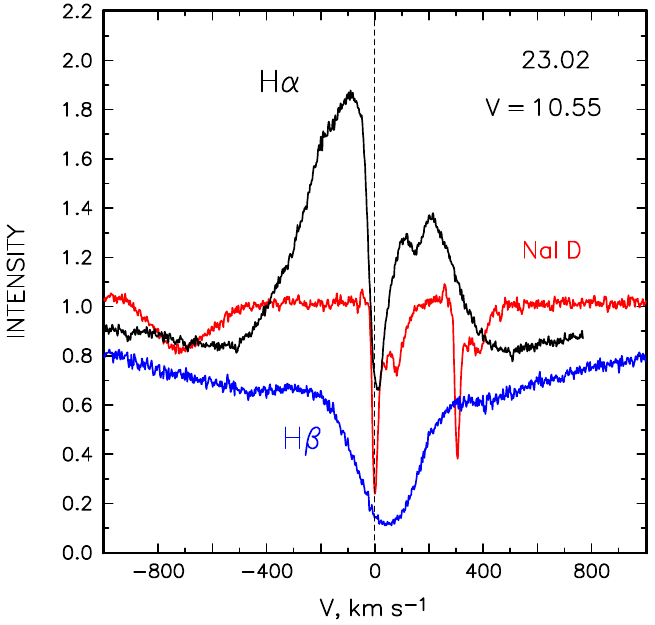}
\includegraphics[width=60mm]{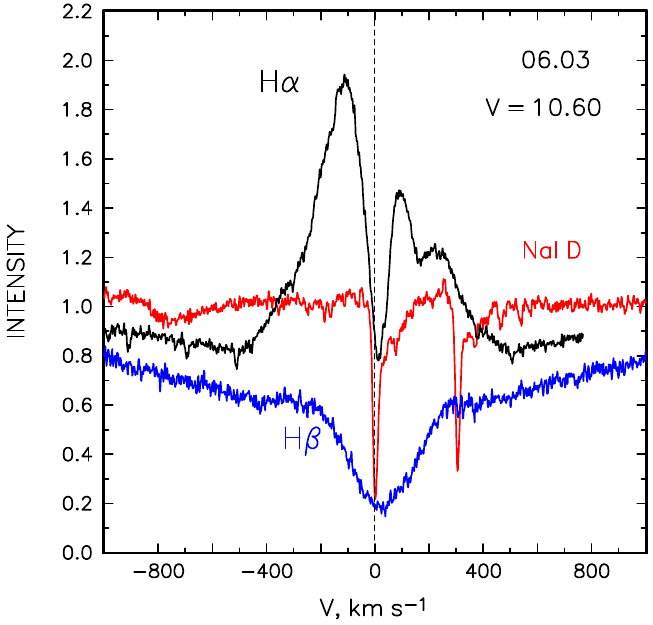}\\
\includegraphics[width=60mm]{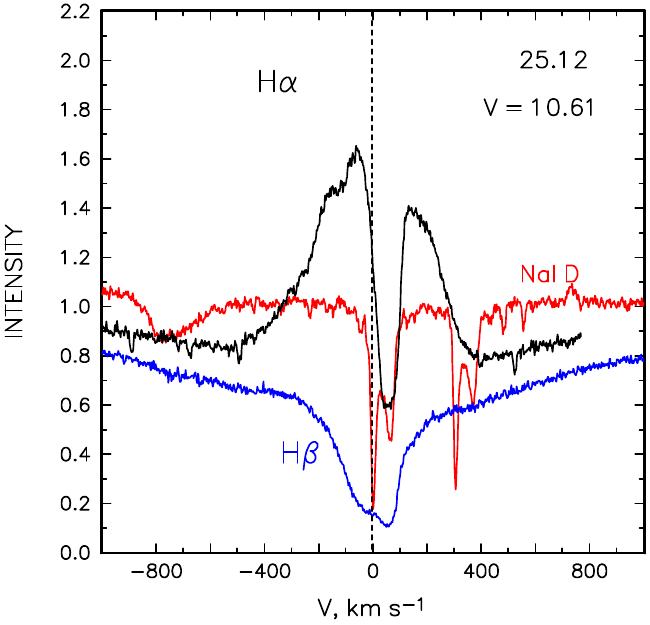}
\includegraphics[width=60mm]{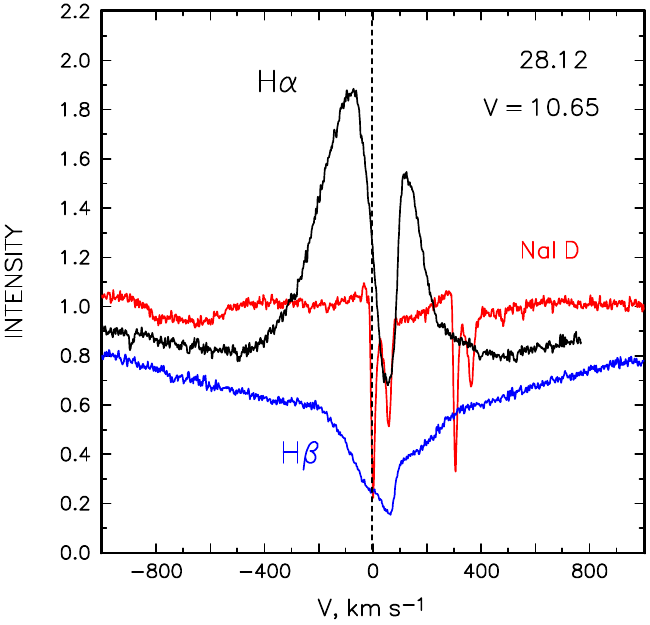}
\includegraphics[width=60mm]{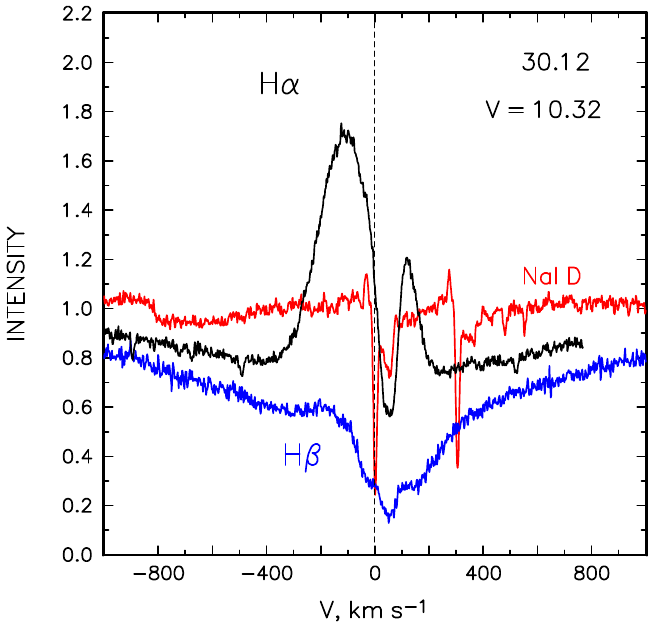}\\
\caption{The line profiles of H$\alpha$ (black), H$\beta$ (blue),
and HeI + NaI~D (red) in 2019 (upper panel) and in 2021 (lower
panel). The central dip in H$\alpha$ is located at about
10~km~s$^{-1}$ in 2019, and it is red-shifted to about
50~km~s$^{-1}$ in 2021. The dates and V-band magnitudes are given
in each panel.} \label{fig7}
\end{figure*}

In December of 2021 we observed that the symmetry of the H$\alpha$
line profile is broken during the whole observational period, see
Fig.~\ref{fig7}. Although the profile remains double-peaked, the
central velocity minimum (dip) is shifted by about 40~km~s$^{-1}$
to the red side, and the "red" emission component is noticeably
narrower. To illustrate this, we have selectively overplotted the
H$\alpha$, H$\beta$ and NaI D$_1$ line profiles obtained on the
same dates for the two periods in 2019 and 2021. All of these
spectra were obtained during the bright state of the star. In
December 2021 the H$\beta$ line has an additional absorption that
appears in the red-shifted absorption profile, and the central dip
of the H$\alpha$ line coincides in velocity with the absorption
component of the sodium line. The sodium doublet forms partially
in the circumstellar medium, in the accretion zone, and partially
in the interstellar medium. The H$\beta$ line forms in a region
that is a few times closer to the star than the H$\alpha$
formation region \citep{gri11,tam20}.

The behaviour of the line profiles shown in Fig.~\ref{fig7} can be
interpreted to suggest that in December 2021 we have caught the
star in a moment of substantial infall of matter, in other words,
an increase in the mass accretion rate. This explanation is
acceptable if we take into account that instability of the line
profiles may be a reaction to the changes in the accretion rate.
This also leads us to suggest that for UX~Ori, the disc wind does
not have enough power to mask the signature of an accretion burst
in the H$\alpha$ line, like it does, for example, in the star
RR~Tau, another member of the UXOR family \citep{grinin2023}. We
also note that during the following dates, beginning of January
2022, the central dip has returned to its usual place, suggesting
the mass accretion rate returned to its former value.

\subsection{Three-peaked H$\alpha$ line profiles}

\begin{figure}
\hspace{-0.4cm}
\includegraphics[width=34mm]{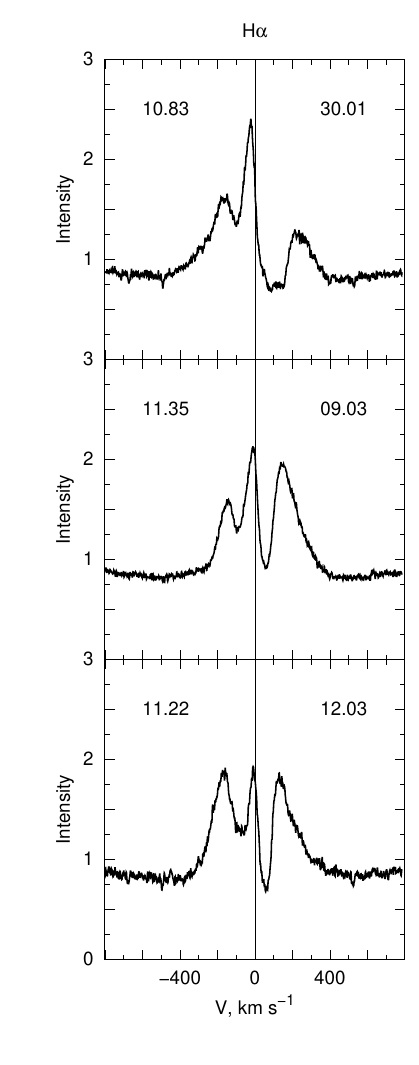}
\hspace{-0.8cm}\includegraphics[width=34mm]{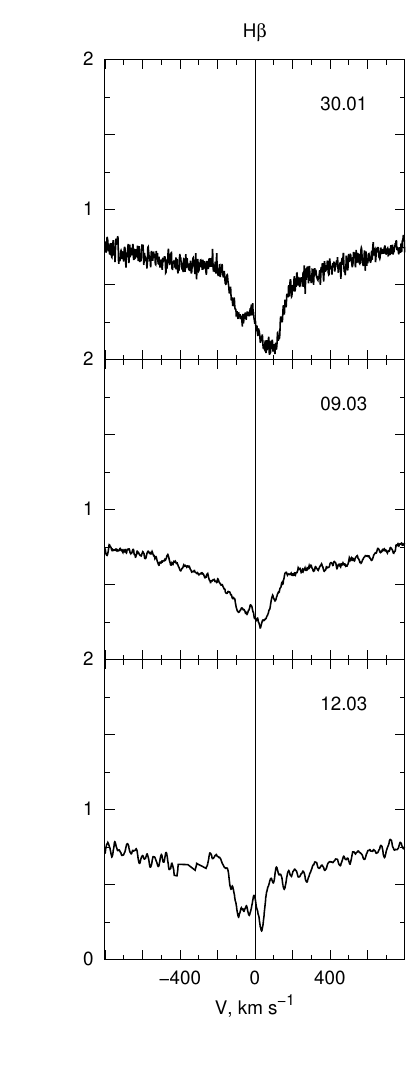}
  \hspace{-0.8cm}\includegraphics[width=34mm]{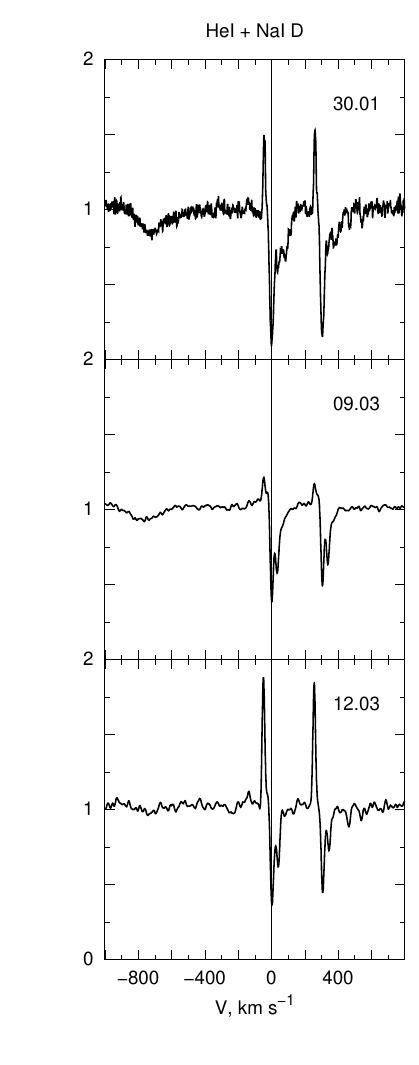}
  \caption{Line profiles of H$\alpha$ (left), H$\beta$ (middle),
    and HeI~5876 \AA \ + NaI~5889/5895 \AA \ (right) in 2020. The dates
    of the observations are shown in each plot, and the left-hand panel
    includes the value of the V-band magnitude.}
  \label{fig8}
\end{figure}

On some dates we observed well-defined three-peaked H$\alpha$ line
profiles (Fig.~\ref{fig8}). All the three emission peaks have deep
gaps between them, and the two blue-shifted emission components
are located at the same positions in all profiles, while the
red-shifted component varies. This gives the impression that we
observe this `three-peak event' as mainly a transformation of the
blue peak, forming an absorption dip in it, and thus being related
to the matter moving towards the observer. We also note that in
the H$\beta$ line profile a small blue-shifted absorption is seen.

The spectrum in the upper panel was obtained when the star was in
the normal bright state, while the two other were obtained on 9
and 12 March 2020, in the fainter state of the star. On March 12
the line HeI 5876 \AA \ disappears, and the sodium lines become
shallower. The latter could possibly be due to the presence of
strong telluric emission lines compared to the weaker ones three
days before, on March 9. We note that the narrow absorption
components of the sodium lines are formed in the circumstellar and
interstellar medium.

We suggest that such a configuration of the profiles may be a
result of evolution of the blue peak due to non-stationary
accretion (i.e. an absorption feature forming in the blue peak).
It could also be explained by the appearance of emission arising
on the periphery of the disc wind that is not obstructed by the
dust screen.

The three-peaked H$\alpha$ line profiles may be one of the
manifestations of a non-stationary accretion. Examples of such
profiles, but with red-shifted lines (in the H$\alpha$) and
blue-shifted absorption components (in the H$\beta$ line) are
shown by \citet[][their Figs. 9 and 10]{bouvier2007}. The authors
considered the young TTS star AA Tau which is seen nearly edge-on.
Investigating the accretion/ejection processes of this star the
authors concluded that the reason of such lines may be the time
variability of the magnetospheric accretion flow on a timescale of
a few rotational periods.

Another explanation is that a third emission peak is formed
between the original blue and red peaks during obscuration of the
star by the dusty screen, where the periphery of the disc wind
remains visible beyond the screen. The Keplerian velocity of this
gas is low, and therefore one can observe a narrow emission at the
low velocities. In order to distinguish between the two possible
explanations, it is required to sample with consecutive
observations the entering into (or going out of) a fading event,
where such a three-peaked profile appears.

\subsection{Equivalent widths of HeI 5876~\AA , NaI~D$_1$, D$_2$, and OI~7774 \AA }

In this section we investigate in more detail the correlation of
the equivalent widths of such lines as HeI~5876~\AA , NaI~D$_2$,
\footnote{D$_1$ is blended with the D$_2$ component and is not
able to give a real information} and OI~7774~\AA. Analysis of the
behaviour of the HeI and OI lines is interesting because their
formation requires high temperatures, because of the high
excitation potential from the lower level, much larger than the
temperature of the star or the disc wind. The latter may be about
of 10000 K for young stars \citep{saf93,garcia01}. Such hot
regions can appear near the star where accreting matter creates
shocks when reaching the stellar surface. When estimating the
equivalent widths (EWs), we took into account only the absorption
part of these lines. We also did not remove their photospheric
components because they are negligibly small \citep[see][Fig. 1
therein]{gri2001}.
\begin{figure}
\centering
\includegraphics[width=9cm]{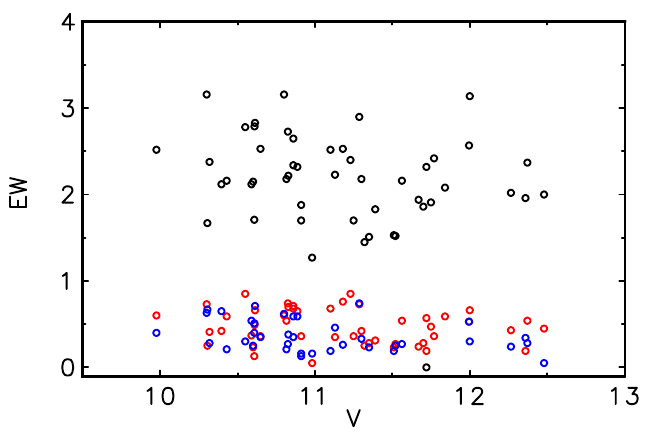}
\caption{Equivalent widths of OI~7774~\AA \ (black), HeI~5876~\AA
\ (red), and NaI~D$_2$ (blue) vs star brightness (V-band
magnitudes) during different observational dates.} \label{fig9}
\end{figure}
\begin{figure*}
\centering \hspace{-1cm}\includegraphics[width=6.2cm]{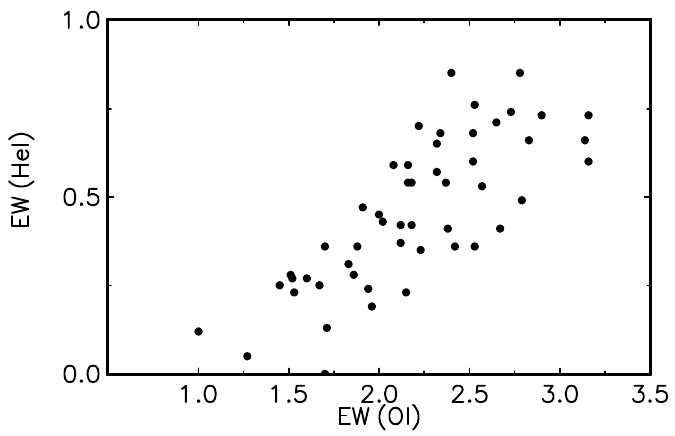}
\hspace{-0.2cm}\includegraphics[width=6.2cm]{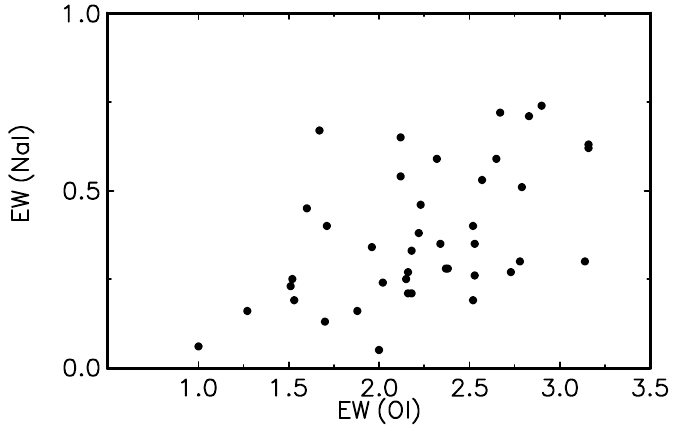}
\hspace{-0.2cm}\includegraphics[width=6.2cm]{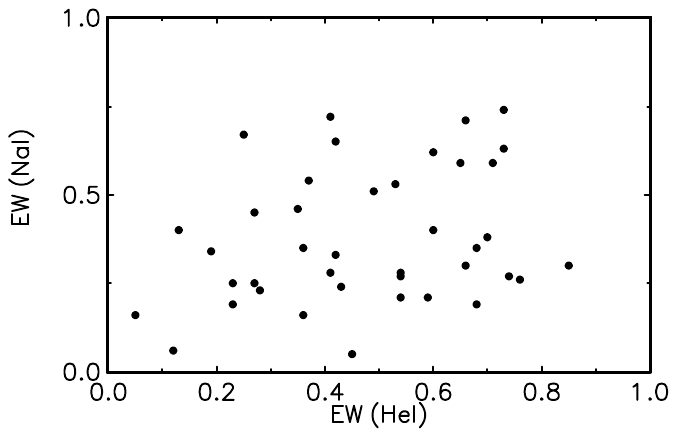}
\caption{Equivalent widths of HeI 5876 \AA \ vs OI 7774 \AA \
(left), NaI D$_2$ vs OI 7774 \AA~ (middle), and  NaI D$_2$ vs. HeI
5876 \AA \ (right). The correlation coefficients are 0.94, 0.64,
and 0.37, respectively.} \label{fig10}
\end{figure*}

Figure~\ref{fig9} demonstrates that the EWs of OI, HeI and NaI do
not depend on the observed V-band magnitude. Nevertheless, the EWs
of HeI and NaI D lines correlate with the EWs of oxygen OI
(Fig.~\ref{fig10}), and the correlation coefficients are 0.94 and
0.64, respectively. The correlation between the equivalent widths
of sodium NaI (D$_2$) and HeI is much weaker, it is 0.37. This
implies that the changes in their line profiles are not caused by
the eclipses of the star and its nearest vicinity. However, this
contradicts the fact that many of the spectra are obtained during
eclipses, i.e. when dust is screening the region of the line
formation partially or fully. We suggest that the behaviour of
these lines may be related to scattered light by the circumstellar
dust. The contribution from scattered light to the stellar flux in
the V-band is substantial, about of 10\% of the radiation of the
star in the bright state and almost 100\% in the deepest minimum
\citep{gri00}.

\section{Similarity and difference in the spectral lines of two stars of the UXOR family}
\label{comparison}

In spite of the fact that UXOR stars belong to the same family due
to their similar evolutionary status and the orientation of the
star--disc system, there may be significant differences in the
main stellar parameters and in the behaviour of their emission
spectra, both in the normal state and during eclipses. We know
from the theoretical study of young stars what physical processes
take place in their environment and in which typical regions these
various processes work \citep[see
e.g.][]{muzerolle2004,kur06,gri11,tam14,brittain2023}. We can
obtain important information from this region by monitoring the
behaviour of main spectral lines in the normal bright state and
the deep minima when the spectra are obtained with the same
telescope/instrument setup. In this section we study how the
spectra of UX~Ori change during the fadings compared to those of
RR Tau, analysed and presented in \citet{grinin2023}. The main
stellar parameters of RR Tau (mass, radius, temperature and $v
\sin i$) are very close to those of UX~Ori. The spectral type of
RR~Tau is A0~III-IV, $\log g = 3.5$, RV~=~+11~km~s$^{-1}$
\citep{gri2001}. The contribution from the scattered light to the
V-band flux is 3\% in the bright state for this star
\citep{gri00}, which is less than that from UX~Ori, estimated
  to about 10\%. We interpret this to indicate that the inclinations
  of the star-disc systems are slightly different for the two stars, since
  the disc properties are similar based on their infrared excesses
  \citep{gri91,natta97}, and thus suggesting that the RR~Tau disc
  is less tilted with respect to the LOS than that of UX~Ori.

\subsection{He~5876~\AA, NaI~D~5889/5895 \AA, and OI~7774~\AA}

\begin{figure*}
\centering
\includegraphics[width=9cm]{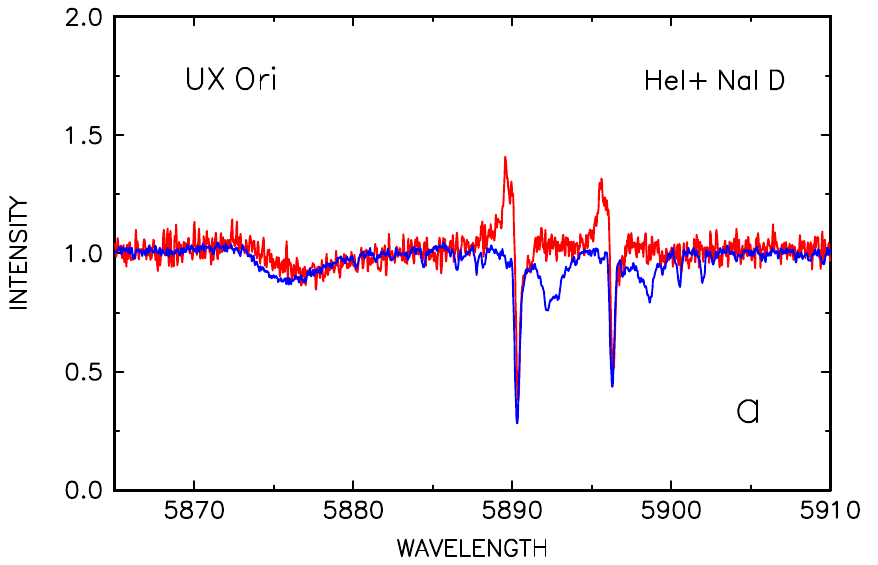}
\includegraphics[width=8.6cm]{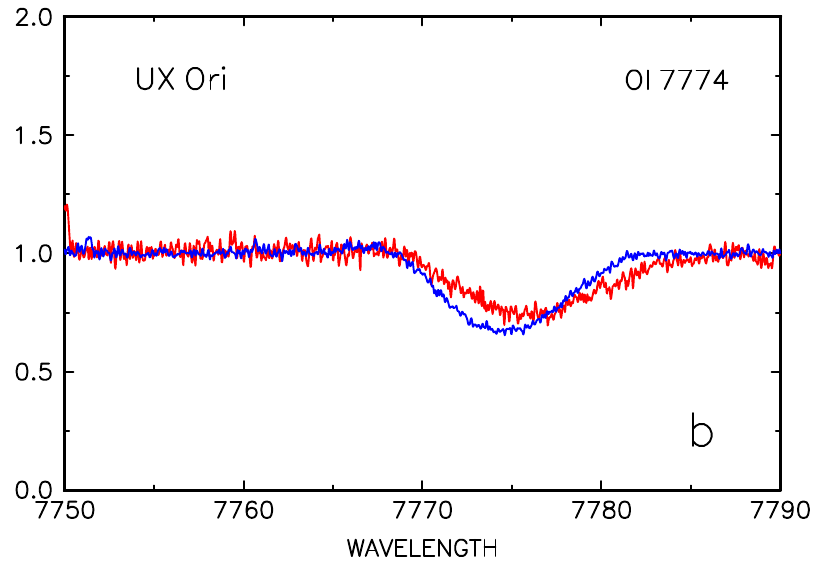}\\
\includegraphics[width=9cm]{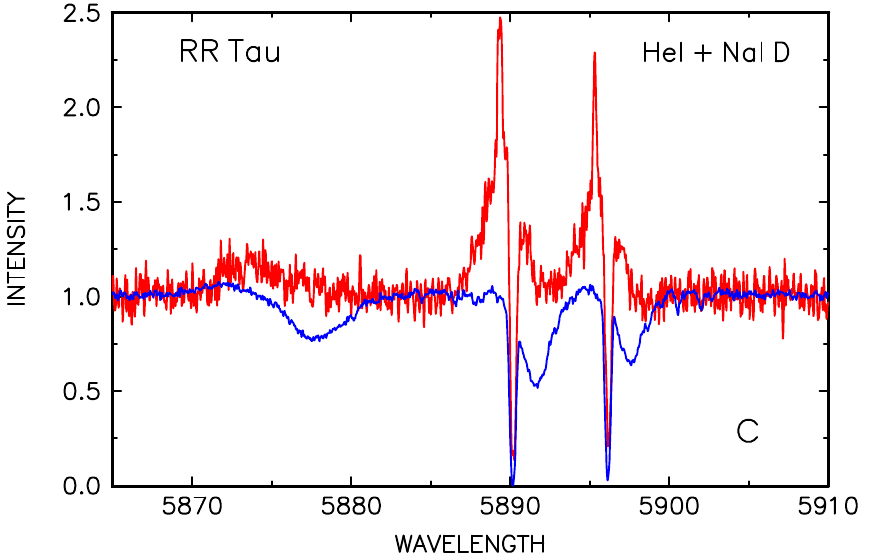}
\hspace{0.1cm}\includegraphics[width=8.6cm]{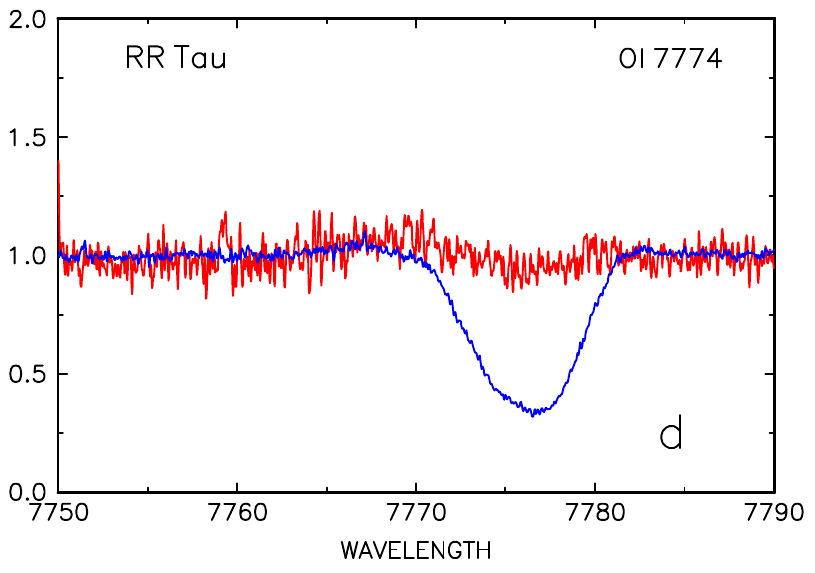}
\caption{\textit{Upper panel (a, b)}: HeI 5876\AA \ + NaI D lines
(left) and OI 7774\AA \ (right) in the spectra of UX Ori in the
normal state on Sep. 5, 2020, V = 9.98 mag (blue) and brightness
minimum on Jan. 18, 2021, V = 12.48 mag (red). \textit{Lower panel
(c, d)}: The same as in the upper panel but for RR Tau: in the
normal state with V = 10.66 mag on Oct 5, 2019, and in the weak
state with V = 13.85 mag on Mar. 15, 2019.} \label{fig11}
\end{figure*}

Let us compare the behaviour of UX Ori and RR Tau during bright
and faint states by looking at the spectral region with the helium
line HeI~5876~\AA~, the sodium NaI~D~5889/5895~\AA \ lines, and
the OI~7774~\AA \ line. Figure \ref{fig11} shows UX Ori in the
upper panels and RR Tau in the lower panels. We have selected two
typical example spectra for both stars during bright states (blue)
and during deep minima (red).\footnote{We mean "deep minimum" if V
$ >$ 12 magnitude (mag) for UX~Ori and V $>$ 13 mag for RR~Tau} In
the case of UX Ori in the bright state, the neutral helium line is
in absorption (panel a), like for RR Tau (panel c), while in the
weak state, it does not disappear or enter in emission, as in the
case of RR Tau \citep{grinin2023}. In UX~Ori only a slightly
red-shifted and shallower line profile is seen, perhaps due to the
appearance of some weak emission. It should be noted that we also
observed a few cases when the HeI line disappears during the
bright state of the star. Thus, we find that for UX~Ori the
disappearance of the HeI absorption line does not correlate with
the star's brightness, and may be instead related to a variable
mass accretion rate. The same may be said about the changes in the
neutral oxygen line (Fig.~\ref{fig11}b,d).

As for the sodium lines, their qualitative behaviour in UX Ori and
RR Tau spectra during bright and weak states are similar.
Nevertheless, it has to be kept in mind that the doublet sodium
lines originate in different parts of the stellar envelope: in the
CS and interstellar (IS) medium unlike the HeI and OI lines.

As mentioned above, the behaviour of the HeI and NaI~D line
profiles in the spectra of UX~Ori do not show an evident
correlation with the star brightness in contrast to what occurs in
the RR~Tau spectra. There were only a few instances when
  the helium line disappeared and the sodium doublet merely consisted
 of the IS component (see e.g. Fig.~\ref{fig3}).
In all those cases the star was in its normal bright state, and
this behaviour may be a consequence of decreasing temperature in
the gas accreting on to the star. During brightness minimum, on
the other hand, the HeI absorption line does not disappear,
probably because the scattered light dominates, since its
contribution is greater for UX~Ori than for RR~Tau, and scattered
light does not make the absorption disappear. This explanation is
not valid for RR Tau because the HeI and OI absorption lines
disappear only during brightness minima when the magnetosphere is
covered by the dust screen, and the contribution from the
scattered light is too small to prevent it.

\subsection{Forbidden oxygen lines [OI] 6300 \AA}
\begin{figure}
\centering
\includegraphics[width=8cm]{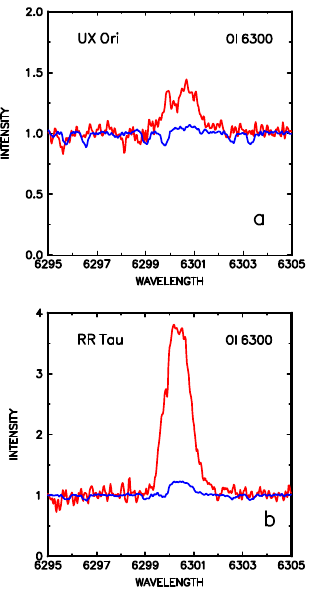}
\caption{Forbidden oxygen lines in the spectra of UX Ori and RR
Tau. \textit{Upper panel (a)}: [OI] 6300 \AA \ lines in the
spectra of UX Ori in the normal state with V = 9.98 mag on Sep. 5,
2020 (blue) and brightness minimum with V = 12.48 mag on Jan. 18,
2021 (red). \textit{Lower panel (b)}: The same as in the upper
panel but for RR~Tau with V = 10.66 mag on Oct. 5, 2019 (blue) and
with V = 13.85 mag on Mar. 15, 2019 (red). } \label{fig12}
\end{figure}

Figure~\ref{fig12} presents  the forbidden lines of the neutral
oxygen [OI] at 6300~\AA \ in the normal and weak states of UX~Ori
(a) and RR~Tau (b). For each star we chose dates of maximum and
minimum brightness. For UX Ori we show the bright state in 2020
(V~=~9.98~mag) and the minimum in 2021 (V~=~12.48~mag).
Correspondingly, for RR~Tau these are in 2019 (V~=~10.66~mag) and
(V~=~13.85~mag). In the normal state of UX~Ori this line reveals a
weak emission but in the brightness minima an additional emission
component appears with a peak ratio blue-to-red (V/R $< 1)$.
However, as an analysis of other line profiles during this minimum
in Jan 2021 and in Jan. 2024 shows that this peak becomes
stronger, its radial velocity remains approximately constant on
the different dates, ranging from -55 to -20 km s$^{-1}$ in the
stellar velocity system. The reason for these almost similar
radial velocities is that all these observations were made in
January, and in the Earth's system all peaks were at the same
velocity, pointing directly to the most likely interpretation that
they originated in the Earth's atmosphere.

\begin{figure}
  \centering
\includegraphics[width=8cm]{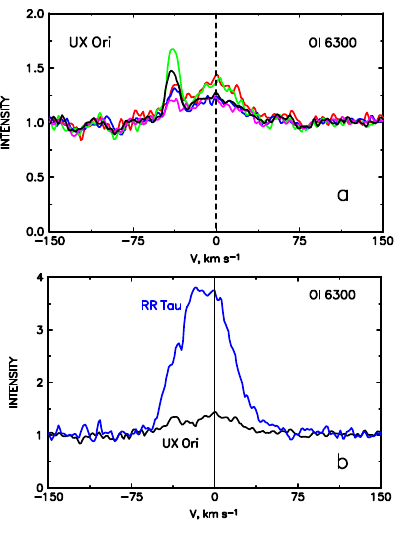}
\caption{Forbidden oxygen line [OI] 6300 \AA \ during brightness
minima. {\em a}: [OI] 6300 \AA \ profiles in the spectra of UX Ori
during the minimum in 2021: Jan. 18, V = 12.48 mag (red); Jan. 20,
V = 12.0 mag (blue); Jan. 24, V = 12.36 mag (green); Jan. 27, V =
12.27 mag (magenta),
  and Jan. 28, V = 12.37 mag (black). {\em b}: Comparison of [OI] 6300 \AA \ profiles in the
 brightness minimum of UX Ori (Jan. 18, 2021, V = 12.48 mag, black) and RR
 Tau (Mar. 15, 2019, V = 13.85 mag, blue). The
arrow points to the telluric lines.} \label{fig13}
\end{figure}

We can conclude that during eclipses, the intensity of the
[OI]~6300~\AA \ line increases and its profiles are slightly
blue-shifted relatively to zero velocity reaching $\sim 75$ km
s$^{-1}$ in the wings. It is seen from Fig.~\ref{fig13} where we
show [OI]~6300~\AA \ line profiles in the deep minima of RR~Tau
(V~=~13.85~mag on Mar. 3, 2019) and UX~Ori (V~=~12.48~mag on Jan.
18, 2021). One can see that in both stars the wings of the
profiles reach the velocity of about $\pm 75$ km s$^{-1}$.

\subsection{FeII multiplet 4924/5018/5069 \AA}

\begin{figure}
\begin{centering}
\includegraphics[width=9cm]{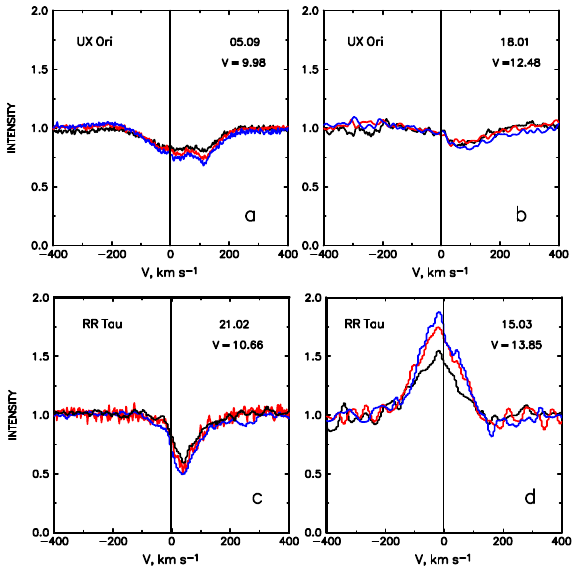}
\caption{Changes in the FeII multiplet in the bright state
and during eclipses. \textit{Upper panel: UX Ori}. The bright
state (left), and the deep minimum (right). \textit{Lower panel:
RR Tau}. The bright state (left), and the deep minimum (right).
Brightness in the visible is shown in each plot. The multiplet
components are (in \AA ): 4924 (black), 5018 (red), and 5069
(blue). The dates of the bright and weak states of stars are the
same as in Fig.~\ref{fig11}} \label{fig14}
\end{centering}
\end{figure}

The behaviour of the FeII multiplet in UX~Ori shown in
Fig.~\ref{fig14}~a,b is analogous to that of oxygen OI~7774~\AA \
and helium HeI~5876~\AA \ in Fig.~\ref{fig11}. In the bright state
of the star these lines exhibit red-shifted absorption profiles
(a), and in the faint state a small
  emission appears at the blue side (b). In the spectra of RR~Tau, these
lines are red-shifted absorption lines in the bright state (see Fig.~\ref{fig14}c)
and blue-shifted emission lines in the faint state of the star. The wings of the
FeII lines reach values of about 200~km~s$^{-1}$ in both stars.

\subsection{SiII 6347/6371 \AA}

Investigation of the silicon lines in the spectra of young stars
is especially interesting because silicon is a part of the
circumstellar dust. Until the dust evaporates, silicon is hardly
found in gas phase, and this usually takes place at the dust
sublimation radius which is at a distance of 0.3 - 0.5 AU from the
star \citep[e.g.][]{tannirkulam2007,flock2017}. In such a case one
can detect a circumstellar 'CS veiling' in the spectra due to
added CS emission. The classical veiling is not related to the
emission in the CS envelope, but to processes on the star itself
(e.g. an appearance of hot spots on the stellar surface).

In the case of the CS veiling, the line is seen not only in
absorption, but also in emission. Therefore, it is interesting to
monitor the behaviour of the SiII 6347/6371 \AA \ lines during the
bright and weak states, and in the latter case the abundance of
gas phase silicon can lead to CS veiling.

Previous observations of UXORs with the NOT \citep{gri2001}, among
them UX~Ori and RR~Tau, gave a rich material, but unfortunately
the stars were in their bright states except for one observation
of RR~Tau. The SiII 6347/6371 \AA \ changed insignificantly in
comparison with other absorption lines. With the current data set
we are in the position to study the SiII 6347/6371 \AA \ line
stability through fading events. We compare the line profiles of
SiII 6347 \AA \ (the strongest of the two) in the bright and weak
states of the stars for UX~Ori and RR~Tau in Fig.~\ref{fig19}. For
reference, in each panel we overplot the profile of the same line
obtained in the normal state of the star. For UX~Ori we find the
following effect: sometimes, in the brightness minima, the SiII
line profile becomes deeper than that in the normal state. We did
not find such an effect in other photospheric lines. Moreover, in
the RR~Tau spectra the silicon line goes in emission during
eclipses. The quantity of this emission may vary, but we did not
find any additional absorption as in the case of UX~Ori. We
suggest that this absorption is not related to mass accretion; but
most likely due to gas located just on the outer boundary of the
magnetosphere where the gas velocity is still low.
\begin{figure}
\begin{centering}
\includegraphics[width=9cm]{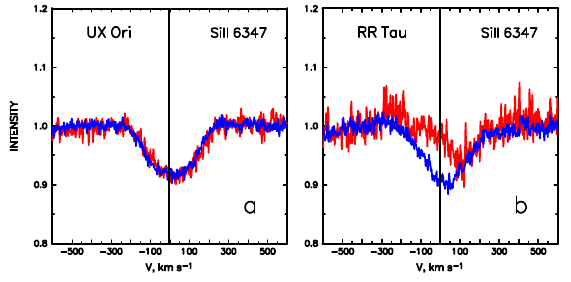}
\caption{SiII 6347 \AA \ line profiles averaged over five
observational dates in the brightness minima (red) and maxima
(blue) for UX~Ori (a) and RR~Tau (b).} \label{fig15}
\end{centering}
\end{figure}

Figure~\ref{fig20} shows other photospheric lines, such as
MgI~5184~\AA \ and FeII~5169~\AA , of UX~Ori during the same period, and for
comparison again we add the line profile obtained in the bright
state of the star. One can see from the figure that the MgI lines
have a very similar shape both in the bright and faint states of
the star, and we note that the FeII line is not deeper during the
minima than in the bright state of the star.

Figure~\ref{fig15} presents the  SiII~6347~\AA \ line profiles
averaged over five different dates during maxima (blue) and minima
(red) of the stars for UX~Ori (a) and RR~Tau (b). It is seen that
in the case of UX~Ori, the average silicon line profiles during
eclipses are similar to those during the normal state. Another
picture is observed for RR~Tau: in the brightness minima
it looks like an inverse P Cygni line profile (red). During the
minima, however, the star and its magnetosphere are eclipsed,
thus, the blue-shifted emission must refer to a wind component.

\subsection{Discrete narrow components}
\begin{figure*}
\centering
\includegraphics[width=90mm]{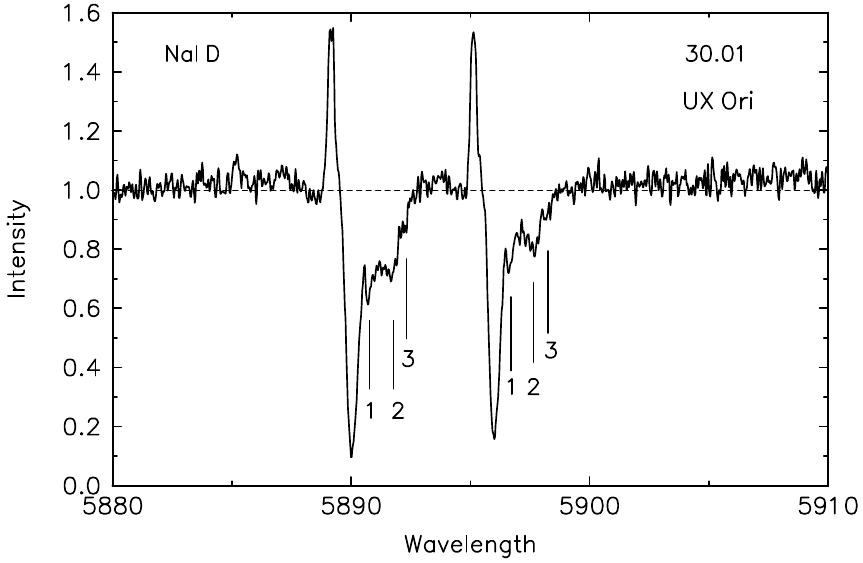}
\includegraphics[width=90mm]{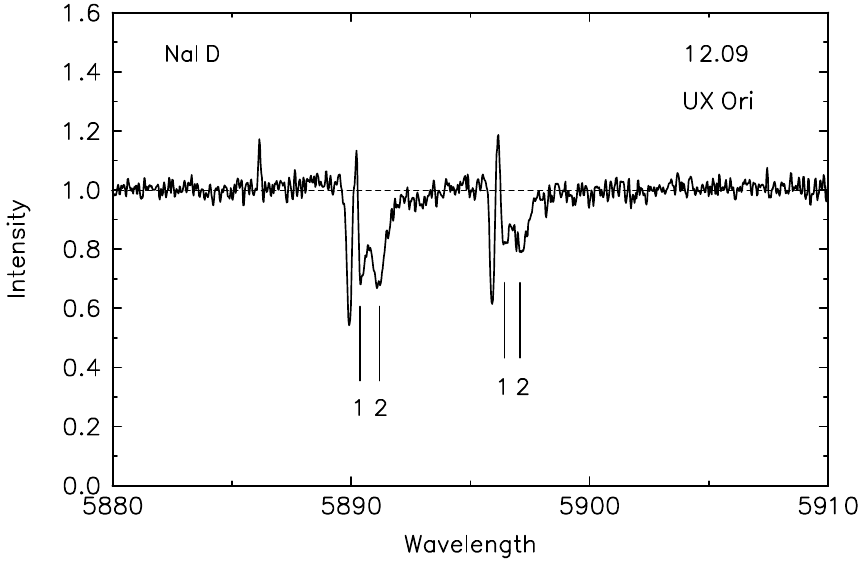}\\
\includegraphics[width=90mm]{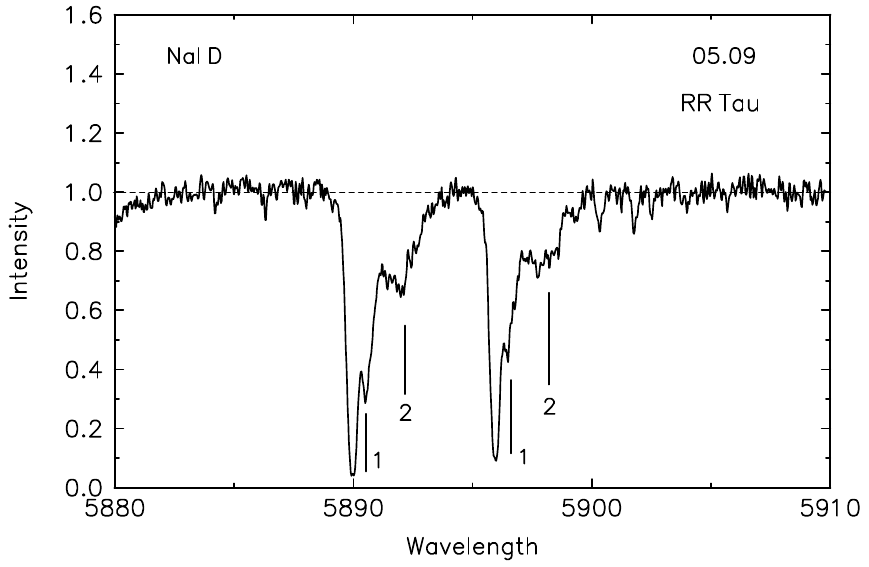}
\includegraphics[width=90mm]{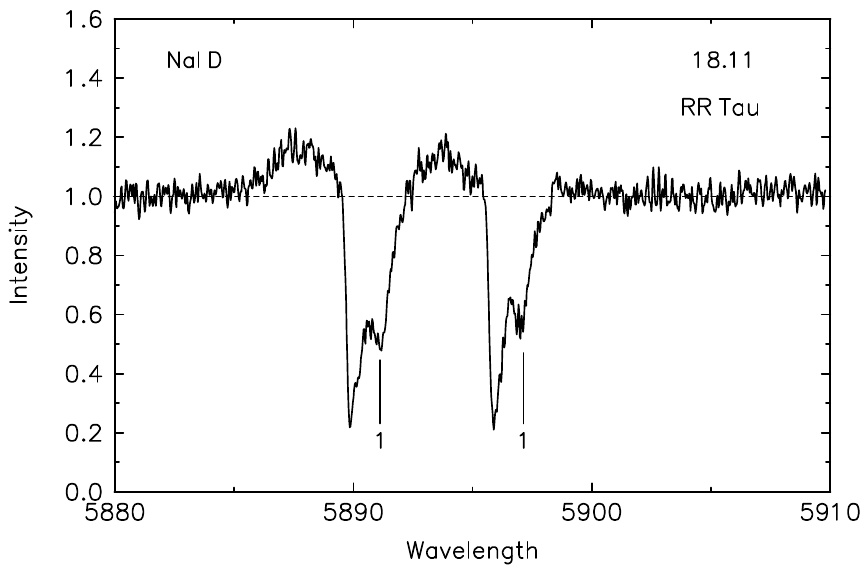}
\caption{Line profiles of the sodium doublet NaI D1 and D2
(5889/5895 \AA) with DACs in the spectra of UX Ori in 2020 (left)
and 2022 (right) (upper panel) and RR Tau in 2019 (left) and 2020
(right) (lower panel). Spectra are shown in the stars' coordinate
systems. See details in the text and in Table~\ref{tab:dac}.}
\label{fig16}
\end{figure*}

Further analysis of the UX~Ori spectra reveal the appearance of
discrete absorption components (DACs) in the sodium line profiles
which are seen when the star is in its normal state.
Figure~\ref{fig16} (upper panel) shows spectra obtained of UX~Ori
on two different dates: Jan.~30, 2020 (V~=~10.83~ mag) and Sep.
12, 2022 (V~=~10.50~ mag). The spectra are presented in the
coordinate system of the star.

The observed DACs are marked in the figure with numbered lines,
and their corresponding velocities in km s$^{-1}$ are shown in
Table~\ref{tab:dac}. Taking into account the uncertainty in the
measurement of the DAC velocities due to the presence of the
telluric lines and noise, we conclude that they are very similar
in both components of the doublet D$_1$ and D$_2$.

\begin{table*}
\caption{Velocities of the discrete absorption components for
UX~Ori and RR~Tau (in km~s$^{-1}$) in the coordinate system of the
star.}
  \begin{tabular}{|c|c|c|c|c|c|c||c|c|c|c|c|}
    \hline
    UX Ori date & D1 1 & D1 2 &D1 3 & D2 1 & D2 2 & D2 3 & RR Tau date & D1 1 & D1 2 &D2 1  &D2 2  \\
    \hline
Jan 30, 2020 & 43 & 94 & 120 & 40 & 96 & 121 & Sep 5, 2020 & 33 & 114 & 34 & 111 \\
\hline
           & D1 1 & D1 2 & D2 1 & D2 2 &  & &             & D1 1 & D2 1 &  &  \\
 \hline
 Sep 12, 2022 & 23 & 64 & 29 & 65 &  &  &     Nov 18, 2019 & 64 & 60 &  &  \\
 \hline

\end{tabular}
\label{tab:dac}
\end{table*}

In the spectra of RR Tau we also found such components in the
NaI~D lines for the two dates: Nov. 18, 2019 and Sep. 5, 2020 as
shown in the lower panel of Fig.~\ref{fig16}. At those dates the
brightness of the star in the visible was 12.54 mag and 11.94 mag,
respectively. We determined the velocities of the absorption
components taking into account the radial velocity of RR Tau. We
consider the discrete character of the red-shifted absorptions in
the NaI~D lines in Sect.~\ref{discussion}.

\section{Discussion}
\label{discussion}

Long-term spectral monitoring of UXORs (2019 - 2024) comprising
both dust eclipses of various degrees, as well as observations of
the stars in their bright states, has given a rich observational material.
In this section we discuss the analysis of several spectral lines of
UX~Ori, and we compare our conclusions for this star with those made
for RR~Tau, another star of the UXOR family \citep{grinin2023}.

\subsection{Hydrogen lines H$\alpha$ and H$\beta$}

The behaviour of the H$\alpha$ emission line during the brightness
minima of UX~Ori is compatible with the model of eclipses where
the dusty screen completely covers the star, but not completely
the more extended line emission region, including part of the
magnetosphere and the disc wind \citep{tam20,grinin2023}. Most of
the emission in this line forms in the disc wind, while the
magnetosphere plays a smaller role. At the moment of eclipses the
intensity of the line increases relatively to the decreasing
continuum. As seen in Figs.~\ref{fig4} and \ref{fig6} the line
profile predominantly becomes almost single-peaked, though in some
cases it retains a double-peak shape. A single-peaked profile as
the one observed in Jan. 2024 and shown in Fig.~\ref{fig6} was
observed only once earlier \citep{gri94}.

Unlike for H$\alpha$, the role of the disc wind is less important
with regards to the formation of the H$\beta$ emission line.  The
main contribution comes from the magnetosphere which is almost
completely obscured by the dust screen during the minima. This is
not surprising since the H$\beta$ line forms in a region which is
closer to the star compared to the H$\alpha$ formation region
\citep[see e.g.][]{gri11,tam20}.

\begin{figure}
\hspace{-0.6cm}
\includegraphics[height=10cm, angle=-90]{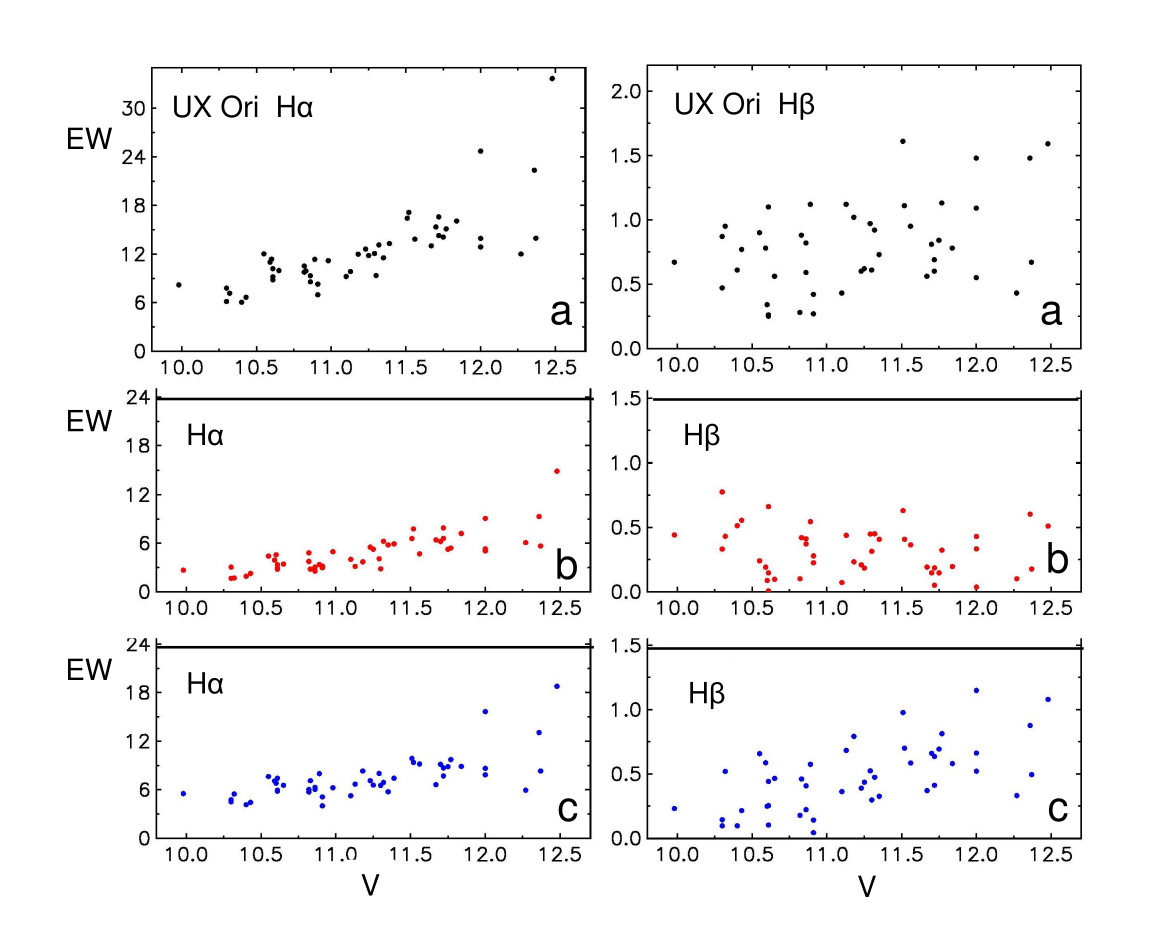}
\caption{Changes in the equivalent widths (EW) of H$\alpha$ (left panel)
  and H$\beta$ (right panel) as a function of the observed brightness of
  UX~Ori given as V-band magnitudes. We distinguish between (a) the total
  EW of the profile, (b) the EW of the red part of the profile where velocities
  are positive,  and (c) the EW of the blue part of the profile where the
  velocity is negative.}
\label{fig17}
\end{figure}

We estimated the equivalent widths of the H$\alpha$ and H$\beta$
lines after having removed the photospheric component by
extracting the synthetic spectrum corrected for rotation.
Figure~\ref{fig17} shows how the equivalent width of the H$\alpha$
line (left panel) increases as the star becomes fainter. As
mentioned above, this is due to the contrast between the emission
line and the continuum. With the three vertical panels we
distinguish between the EWs of (a) the total line, (b) the red
part of the double-peaked profile belonging to the positive radial
velocities, and (c) the blue part with negative radial velocities.
It is seen that the EWs of both the red and the blue components of
the H$\alpha$ line profiles increase when the brightness
decreases. Observations of the RR~Tau spectra \citep{rod02} lead
to the same conclusion. The behaviour of H$\beta$, however, shown
in the right hand panels, does not have the same trend. While the
blue component clearly increases when the star fades, the red
component is more scattered with no clear trend.

\begin{figure}
\includegraphics[width=9cm]{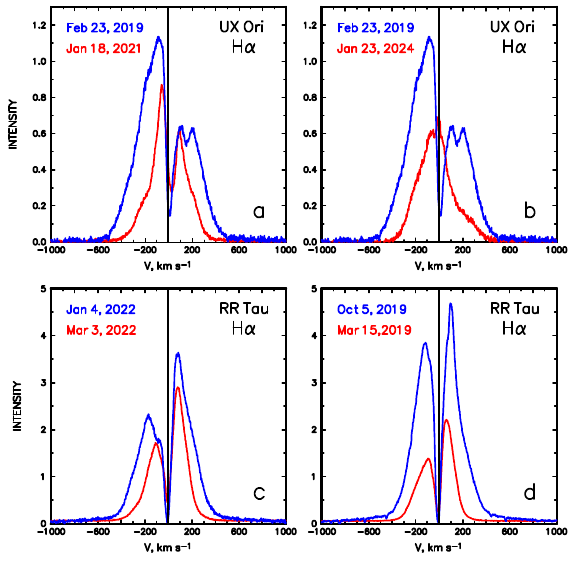}
\caption{Comparison of the H$\alpha$ emission line profiles,
scaled to the same energetic units in the bright and faint states.
\textit{Upper panel:} The H$\alpha$ emission line profiles in the
normal bright state (blue) at V = 10.55 mag and two different
faint states (red) of UX~Ori. The epochs of each profile are given
in the plots. In the two faint states the photometric measurement
are a) V = 12.48 mag and b) V = 11.70 mag. \textit{Lower panel}:
Same as in the upper panel, but for bright and faint states of RR
Tau observed in 2019 (c) and 2022 (d). The photometric measurement
in the bright and faint states are V = 10.86 mag and V = 13.77 mag
(c), and V = 10.66 mag and V = 13.85 mag (d); see text for
details.} \label{fig18}
\end{figure}

We compared the emission profiles of several H$\alpha$ lines
obtained in the normal state of the star and during the minima.
For this we shifted the observed spectra to the star's coordinate
system in the region of the line and subtracted a synthetic
spectrum. After this we reduced the chosen profiles to the same
energetic units. In order to determine the reduction coefficient
we calculated the radiation flux in the continuum $F_i$ at the
observational moment $i$ as
\begin{equation}\label{e12}
    F_i=10^{-0.4 *\Delta V},
\end{equation}
where $\Delta V = V_i-V_0$, and $V_0$ is the star's magnitude
taken in the bright state. The value of $V_0$ was chosen by
averaging over the five magnitudes obtained in the bright states,
getting a mean V$_0$ = 10.43 mag. Then, the emission spectra of
all chosen lines, including that taken in the bright state, were
multiplied by the reduction coefficients defined from the above
relation. The comparisons are shown in Fig.~\ref{fig18} in panels
a and b. The same procedure was applied to spectra of RR~Tau
during bright and faint states (panels c and d).

We see that during eclipses the H$\alpha$ line profiles become
narrower than in the bright states. This takes place because
during the brightness minima the dust screen covers the star and
the region near the star where the Keplerian velocity is the
highest. Depending on the position of the screen, we can see the
line profiles either nearly symmetric around the central velocity
or slightly asymmetric (Fig.~\ref{fig18} b). Not only the position
of the screen, but also the non-homogeneous structure of the
emitting region may cause such an asymmetry.

\subsection{Special features of the line profiles. Comparison with RR~Tau}
\label{special-features}

Comparative analysis of the spectral variability of UX~Ori and
RR~Tau reveal the following differences: the H$\alpha$ line in the
spectra of UX~Ori demonstrate a greater variety of profiles than
RR~Tau, including the above mentioned three-peaked profiles that
never were observed in the spectra of RR~Tau, but are episodically
observed in the spectra of T\,Tauri stars \citep{bouvier2007}.
This suggests that the emission region of UX~Ori may differ from
that of RR~Tau. Our analysis permits us to conclude that there is
a different ratio between the power of the accretion and disc wind
(i.e. the disc wind of UX~Ori is weaker than that of RR~Tau). This
is consistent with the weaker mass accretion rate onto UX Ori
compared to that onto RR Tau given by \citet{guzman2021}.

Let us compare the influence of the accretion process on the line
profiles. The H$\alpha$ profile is typical for the disc wind
structure, but its asymmetry (the peak ratio V/R is always greater
than one) is a sign of accretion (see e.g. Fig.~\ref{fig2}). Such
a profile may be obtained if the emission in the red component is
reduced due to absorption by the accreting gas. The H$\beta$
line profile supports this interpretation because its asymmetry
remains in the inverse P~Cygni shape. \citet{natta-grinin2000}
show spectra of UX~Ori where the absorption features in the red
wing of the H$\alpha$ line correspond to the location of the
absorption features in the sodium lines (see Fig. 2 in their
paper). This fact was used by the authors as an argument for the
presence of magnetospheric accretion. In the present observations
the absorptions in the red components of the NaI~D lines is also
at about the same velocities as in the H$\alpha$ lines (see
Fig.~\ref{fig7}).

When comparing changes in the different line profiles during the
eclipses, we note several differences between UX~Ori and RR~Tau.
For RR~Tau the H$\alpha$ line shape stays double-peaked during
eclipses, while for UX~Ori the profile may stay double-peaked or
become single-peaked. In the case of RR~Tau the H$\beta$ profile
takes the shape of inverse P~Cygni during bright states and the
double-peaked emission profile with changing V/R ratios during
eclipses. In UX~Ori, on the other hand, we find that during
brightness minima an absorption feature appears bluewards of the
inverse P~Cygni H$\beta$ line shape. At the line centre of
H$\beta$ an additional emission feature may appear (see e.g.
Fig.~\ref{fig6}).

In RR~Tau the photospheric lines which are in absorption during
the bright states transform into the emission lines during the
minima, while in UX~Ori only CaII~8642~\AA \ enters in emission
during the minima. Other lines either show presence of a weak
emission (the FeII triplet) or practically do not change (e.g.
SiII~6347, 6371~\AA , and MgI~5184~\AA ).

We did not find a clear correlation between absorption on the red
side of the NaI~D profile and the star's brightness. The reason
may be related to the fact that these lines form in different
regions, which makes the analysis difficult. The NaI~D line
partially forms in the accreting region and partially on the CS or
IS periphery. The discrete character of the red-shifted
absorptions in the NaI~D lines in Fig.~\ref{fig16} testifies to
the extremely irregular distribution of the accreting matter in
the magnetosphere of UX~Ori and to the non-stationary character of
the accretion. We observed the DACs in both stars; their
  appearance may be explained by either a gas stream that
  accretes on to the star and circles the star several times with the
  different infall velocities, or a non-homogeneous gas (blob) which
  accretes directly onto the star.

  \begin{figure*}
\begin{centering}
\includegraphics[width=13cm]{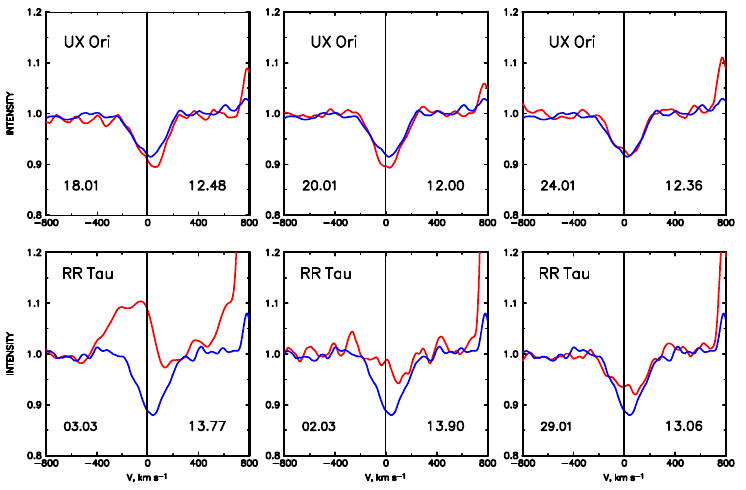}
\caption{Si II 6347 \AA \ in the spectra of UX Ori in 2021 (upper
panel) and RR Tau in 2022 (lower panel, left) and in 2021 (lower
panel, middle and right) during the brightness minima (red). The
dates of the observations and V-band magnitudes are shown in each
plot. The blue lines represents the bright state reference SiII
line profile. Shown are the observed profiles in the bright states
on Feb. 20, 2019, with V = 10.59 mag for UX~Ori, and on Jan. 7,
2022, with V = 10.58 mag for RR~Tau.} \label{fig19}
\end{centering}
\end{figure*}

\subsection{Lines of the high-excitation HeI 5876 \AA \ and OI 7774 \AA}

Very interesting differences are shown by the two absorption lines
HeI 5876 \AA \ and OI 7774 \AA. Both of them originate in the hot
gas of the stellar magnetosphere, because only there one can
expect sufficiently high temperatures for the excitation of these
lines.

These absorption lines disappear during the brightness minima of
RR~Tau due to filled in emission which is not seen in the normal
bright state. In the spectra of UX~Ori, on the other hand these
absorption lines do not disappear, they become weaker but remain
visible, as shown in Fig.~\ref{fig11}. We propose that these
differences are caused by the different level of scattered light
from the protoplanetary disc.  In the case of UX~Ori the
contribution from scattered light is noticeably higher than in the
case of RR~Tau. This follows from the differences in the
photometric amplitudes at the minima, which are $\Delta$V = 2-3
magnitudes for UX~Ori and $\Delta$V =  3-4 magnitudes for RR~Tau.
During eclipses the relative contribution from scattered light to
the total brightness of the star increases, and the amount of
scattered light sets a limit to how large the amplitude at minimum
can be. In the bright state the scattered light contributes 10\%
of the observed radiation in the case of UX~Ori, and only 3-4\% of
the observed radiation in the case of RR Tau.

From the analysis of the UX Ori spectral lines in the
  bright state (H$\alpha$, HeI 5876 \AA , NaI~D and OI~7774~\AA )
  we conclude that the gas falls on to the star with a velocity of
  about 200~km~s$^{-1}$. The outflow velocity is about of
  400~km~s$^{-1}$. Approximately the same values are valid for
  RR~Tau. A detailed comparative analysis of parameters and features of emitting
regions of these stars will be given after modelling of the UX~Ori
hydrogen line profiles in Part~II.

\subsection{Photospheric lines}
\label{discussion-photospheric}

Investigation of photospheric lines (e.g. Fe, Si, Mg) plays an
important role in our understanding of the physical and
geometrical structure of the regions nearest to the star, and the
processes taking place there, because they are not expected to be
with CS absorption or emission lines. The photospheric lines are
used when defining important parameters (e.g. the rotational
velocity of the star, $\log g$).

Nevertheless, we observe that photospheric lines show line profile
changes during brightness minima due to the appearance of emission
that is hard or impossible to see in the bright state. During
brightness minima one can determine the possible place of the line
formation and its intensity as we have done in the case of RR~Tau
\citep{grinin2023}. In the spectra of RR~Tau the appearance of the
emission is clearly seen, while in UX~Ori these line changes are
less clear (see Figs.~\ref{fig14}, \ref{fig19}, and \ref{fig20}).
For both stars we interpret the character and source of the
additional emission to be that part of the disc wind which arises
on the periphery of the CS disc and is not obscured during an
eclipse. One can expect the appearance of CS veiling that is
caused by processes occurring in the CS envelope.

It should be noted that the scattered radiation of the
  protoplanetary disc can broaden the photospheric lines of  T~Tauri
  stars but only weakly influence the photospheric lines of UXORs.
  Both UX~Ori and RR~Tau are rapidly rotating stars and for them
  broadening of these lines is negligible \citep{gri22}. However, the
  radiation of the young stars may be scattered, not only in the dusty
  disc, but also in the magneto-centrifugal disc wind. The latter lifts
  small dust particles \citep{saf93}. Therefore, the disc wind
  may be an additional source of scattered radiation. In this case,
  the photospheric lines are both broadened due to a differential
  rotation of the disc wind and possibly red-shifted \citep{gri06,gri12}.

\subsection{Forbidden oxygen line [OI] 6300 \AA}
\label{discussion-oxygen}

We compared the behaviour of the forbidden oxygen lines in the
RR~Tau and UX~Ori spectra in Fig.~\ref{fig13} and concluded that
the changes in line profiles are identical, and differences are
only in the scale of the processes. The line profiles have the
following structure: they contain a low-intensity emission profile
slightly blue-shifted relatively to zero velocity. The place of
the formation of the low-intensity emission in UX~Ori is identical
to that in RR Tau: these are regions of low density where matter
outflow exists but rotation is small. Such characteristics fit
well with the region where the photoevaporative disc wind forms,
in the periphery of the protoplanetary disc \citep[see e.g.][and
references
    therein]{erco10,erco16, mcgin18,bal20,pasc23}. The line
profiles are symmetric relatively to zero velocity, the wind
velocity reaches $\sim 75$~km~s$^{-1}$ that coincides with the
velocity of the same line in the RR Tau spectra.

\subsection{Summary data on the line formation regions and line
properties for both stars} \label{App-properties}

Having analysed the lines in the spectra of UX~Ori and RR~Tau, we
have highlighted the main conclusions about their properties and
regions of their formation as well as their similarity and
difference.

The H$\alpha$ line forms in the circumstellar (CS) region in the
accretion and disc wind zones. In both stars the H$\alpha$ line
profiles demonstrate the double-peaked profile in the bright state
with a deep gap at the velocity close to zero, with the
blue-to-read peak ratio V/R $>1$ (UX Ori) and V/R $\leq1$
(RR~Tau). During eclipses the emission lines grow relatively to
the decreasing continuum by a factor of 1.5 - 3 (UX~Ori) and a
factor of about 8 (RR~Tau) compared to their bright state. During
the brightness minima the UX~Ori line profiles can keep their
double-peaked shape, but more often they transform into a
single-peaked profile, while for RR~Tau the double-peaked line
profiles are also retained during minima.

The H$\beta$ line forms in the CS envelope (in accretion
and disc wind zones) in a smaller area compared to that of the
H$\alpha$ line formation. In the bright state its line profiles
are in absorption and a little red-shifted (UX~Ori) and have an
inverse P~Cygni shape (RR~Tau). In the brightness minima of UX~Ori a
weak emission is seen at the blue side: the line profiles become
slightly wider and shallower. In the brightness minima of RR~Tau
the line profiles become intense and double-peaked with the wind
features: V/R $< 1$.

The HeI~5876~\AA \ and OI~7774~\AA \ form in the accretion
region of CS envelope of both stars. In the bright state their
line profiles are in absorption. In the brightness minima of UX~Ori
they do not change.  In the brightness minima of RR~Tau the
absorption disappears.

The sodium doublet NaI~D lines originate in the accretion
part of the CS region and partially in the interstellar medium. In
both stars they are in absorption. In the brightness minima of
RR~Tau the emission can appear while in that of UX~Ori the line
profiles remain in absorption.

The lines of the CaII (8498/8542/8662 \AA ) and FeII
(4924/5018/5069 \AA ) triplets and SiII (6347/6371 \AA ) doublet
originate in the star and the accretion part of the CS envelope.
In the bright state of both stars the CaII and FeII lines are in
absorption and red-shifted, resembling the inverse P~Cygni profile.
In the brightness minima the wind emission is added from the blue
side: the profiles remain in absorption (UX~Ori) and transform
into the emission ones (RR~Tau). The SiII lines are in absorption
in the bright state of both stars. They are almost symmetric
relatively to the zero velocity (UX Ori) and they are red-shifted
(RR Tau). In the brightness minima they did not change (UX Ori)
and transforms to the profiles with the blue - side emission (RR
Tau).

The forbidden oxygen [OI] (6300/6363 \AA) lines have a
single-peaked profile in the bright state for both stars, in the
brightness minima the profiles keep their shape and their
intensity increases weakly (UX Ori) and significantly (RR Tau).
The luminosity of the lines remains constant during eclipses, this
means that they are formed in a region beyond the dust screen. The
line profiles are practically symmetric relative to zero velocity
(UX Ori) and slightly blue-shifted (RR Tau). The source of their
formation may be a photoevaporated disc wind on the periphery of
the disc (both stars) or a weak jet manifesting itself only
spectrally (RR Tau).

We can conclude more quantitatively about the regions of
  the hydrogen line formation and thus on the accretion and mass loss
  rate after modelling of the eclipses in UX~Ori.

\section{Conclusion}

Spectral lines of young stars observed through the atmospheres of
their protoplanetary discs, due to their specific orientation,
give important information about the physical processes occurring
in the vicinity of the star. We analysed the spectra of UX~Ori
obtained during the long-lasting (five years) period of monitoring
and Target-of-Opportunity observations with the NOT, and compared
them with those obtained for RR~Tau with the same telescope in the
same period. We have made the following conclusions:

\begin{itemize}
\item During eclipses of the star by the gas and dust fragments
acting as a screen, a strengthening of the H$\alpha$ emission line
relative to the decreasing continuum occurs. At the same time the
line profile becomes narrower because the opaque screen covers the
star, and also that part of the moving gas of the H$\alpha$
emitting regions where the Keplerian velocity is the highest.
During eclipses the H$\alpha$ line
profiles may be double-peaked or single-peaked. \\

\item During eclipses additional emission appears at the
frequencies of the photospheric lines (FeII, CaII, SiII and
others). This may lead to circumstellar veiling of these lines. \\

\item When determining fundamental parameters of UXORs (e.g.
$T_{\rm eff}$, $\log$ g, v $\sin$ i) one has to use the spectra
obtained in the bright state of the stars because in brightness
minima the emission component of the disc wind lines that are not
obscured by the dust screen is added to the photospheric lines.
For the same reason, the radial velocity measurements of these
stars may be red-shifted.\\

\item The two UXOR stars UX~Ori and RR~Tau display different
behaviours in some of the spectral lines during the brightness
minima. The reason for these differences may be either a different
contribution of the scattered light component to the total stellar
radiation during eclipses and/or a less
  intense disc wind in the case of UX~Ori. \\

\item Our observations showed that in the spectra of UX Ori and of
RR Tau the diffusion interstellar band (DIB) at 6283~\AA \ did not
change, even when the brightness decreased significantly. This
means that the CS extinction does not add anything to the DIB and
that the DIB has a purely interstellar origin.

\end{itemize}

Because of the specific orientation of the star--disc system of
young stars of the UX Orionis family, observations of these stars
permits us to study details in the spectra that cannot be seen in
other young stars. In this paper we have presented long-term
spectral monitoring of UX~Ori, and we have compared the results to
those of RR~Tau. Further spectroscopic studies of this star type
during their brightness minima will give information about the
location and structure of the emitting regions near the young
stars due to the enhanced contrast between the line emission and
the star continuum provided by the eclipses.

\begin{acknowledgements}
We thank the referee for helpful comments and suggestions
concerning this work.
  Based on observations made with the Nordic Optical Telescope, owned in collaboration
  by the University of Turku and Aarhus University, and operated jointly by Aarhus
  University, the University of Turku and the University of Oslo, representing Denmark,
  Finland and Norway, the University of Iceland and Stockholm University at the Observatorio
  del Roque de los Muchachos, La Palma, Spain, of the Instituto de Astrofisica de Canarias.
  AAD would like to thank all the students and staff at the NOT for all the work in making
  the flexible scheduling possible and efficient. A special thanks is directed to John Telting
  for supporting and maintaining the FIEStool pipeline. A.A.D. and H.W. would like to thank
  Jens Hoeijmakers and Nicholas Borsato for advice and help related to the installation and
  use of the {\em molecfit} program for FIES data.  V.P.G. and L.V.T. would like to acknowledge
  the support of the Ministry of Science and Higher Education of the Russian Federation under
  the grant 075-15-2020-780 (N13.1902.21.0039).

\end{acknowledgements}

\section{Data Availability}
All the FIES data used in this paper are available in the Nordic
Optical Telescope online FITS
archive.\footnote{\url{http://www.not.iac.es/obsering/forms/fitsarchive/}}
The FIEStool pipeline reduced spectra are also available in the
NOT archive. Practically all the photometry observed by SAAF is
available in AAVSO\footnote{\url{https://www.aavso.org/}} or
otherwise available by request to the authors.

\bibliographystyle{aa}
\bibliography{uxori1}
\begin{appendix}
\section{FIES/NOT spectra of UX Ori}
\begin{figure*}
\hspace{-0.4cm}\includegraphics[width=36mm]{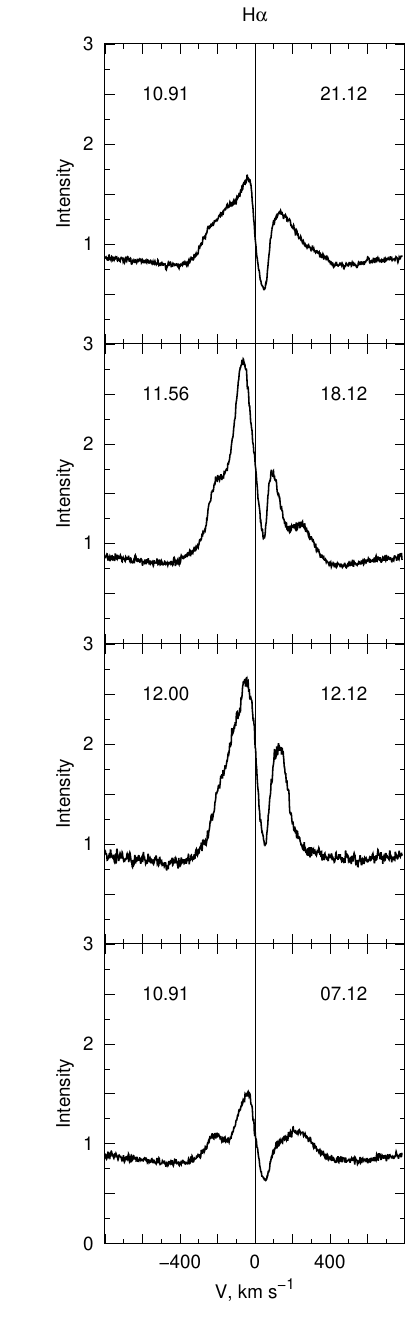}
\hspace{-0.6cm}\includegraphics[width=36mm]{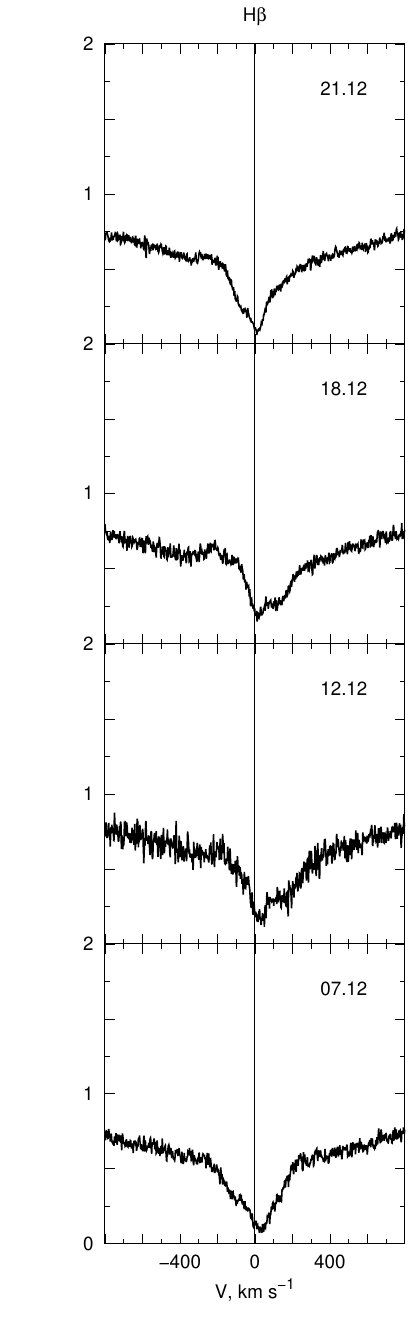}
\hspace{-0.6cm}\includegraphics[width=36mm]{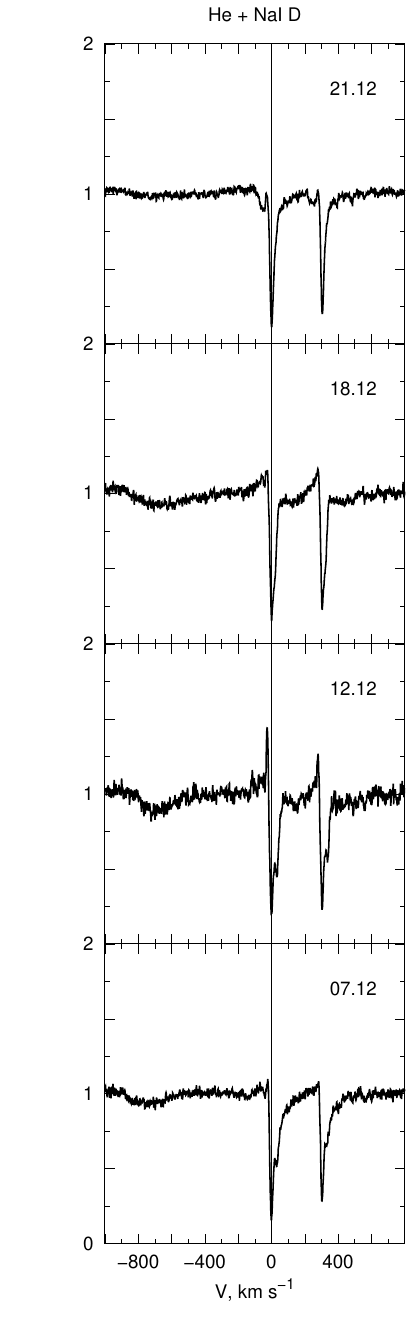}
\hspace{-0.6cm}\includegraphics[width=36mm]{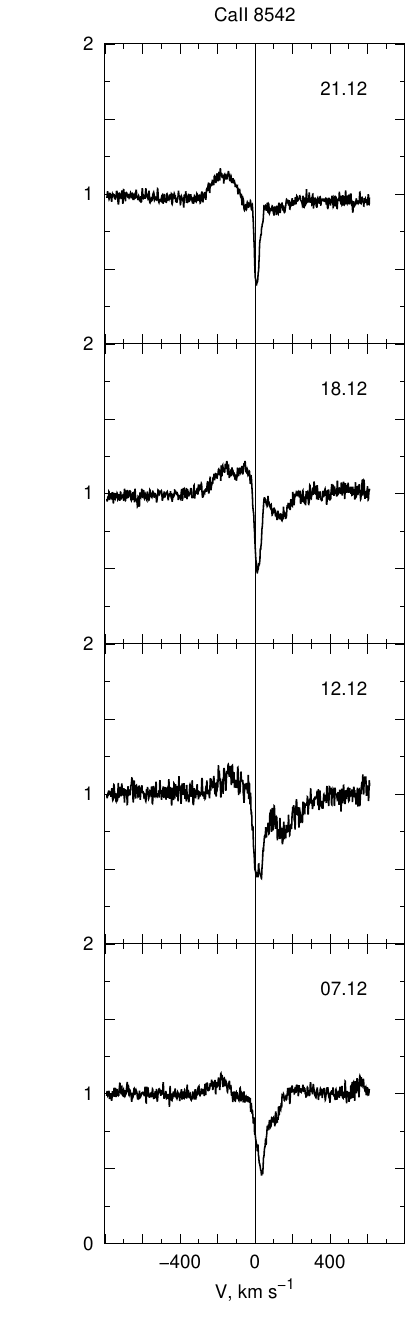}
\hspace{-0.6cm}\includegraphics[width=36mm]{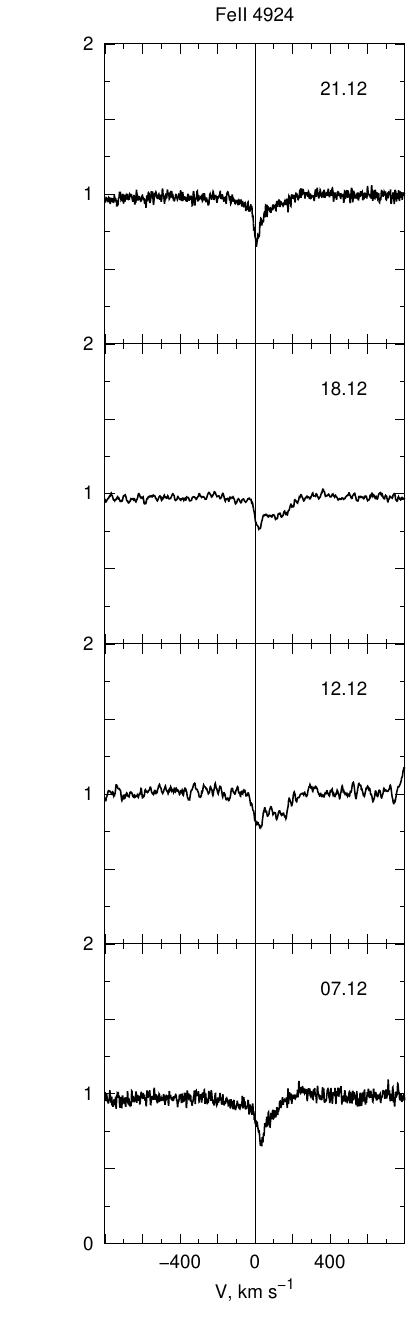}
\caption{The same as in Fig.~\ref{fig2}, but when entering the brightness minimum in December 2020.}
\label{fig3}
\end{figure*}

\begin{figure*}
\hspace{-0.4cm}\includegraphics[width=36mm]{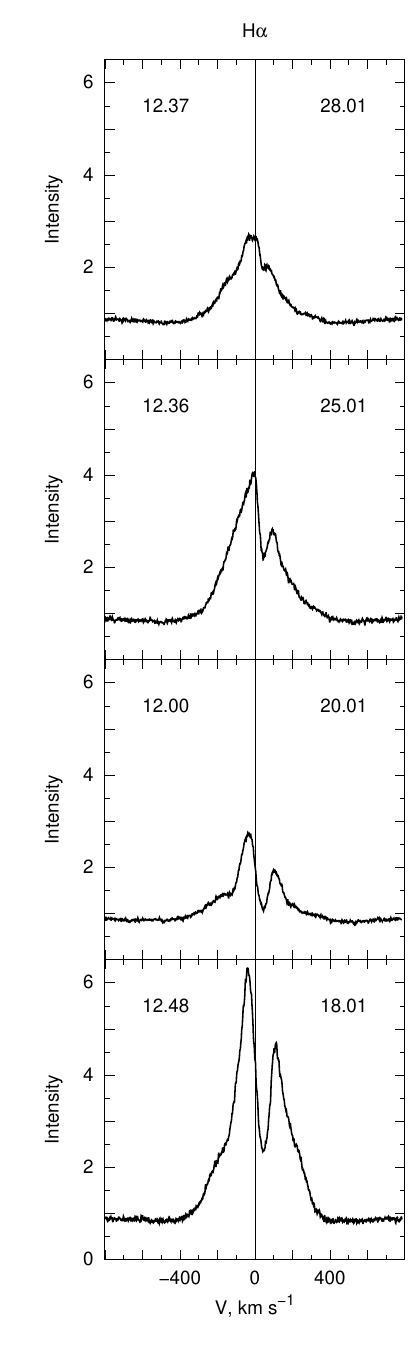}
\hspace{-0.6cm}\includegraphics[width=36mm]{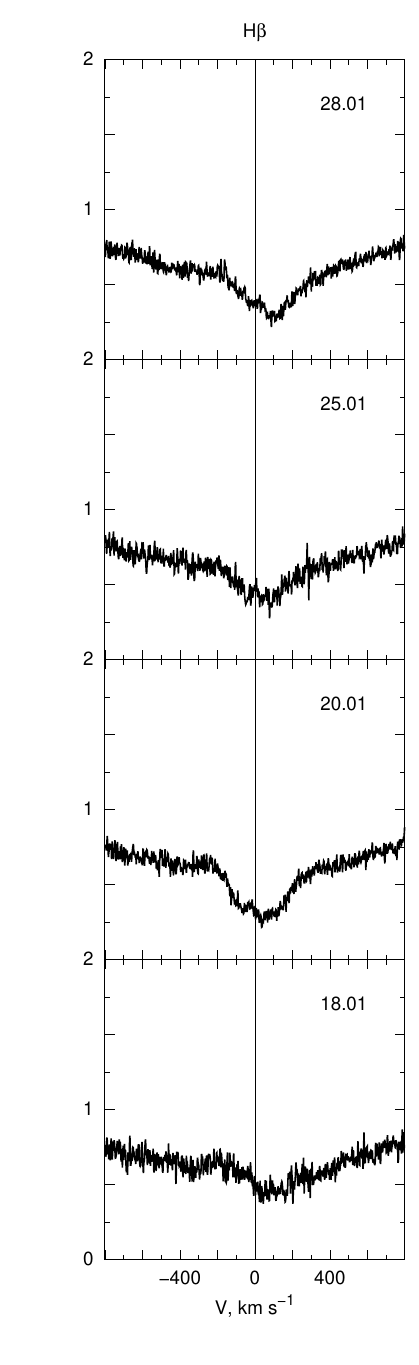}
\hspace{-0.6cm}\includegraphics[width=36mm]{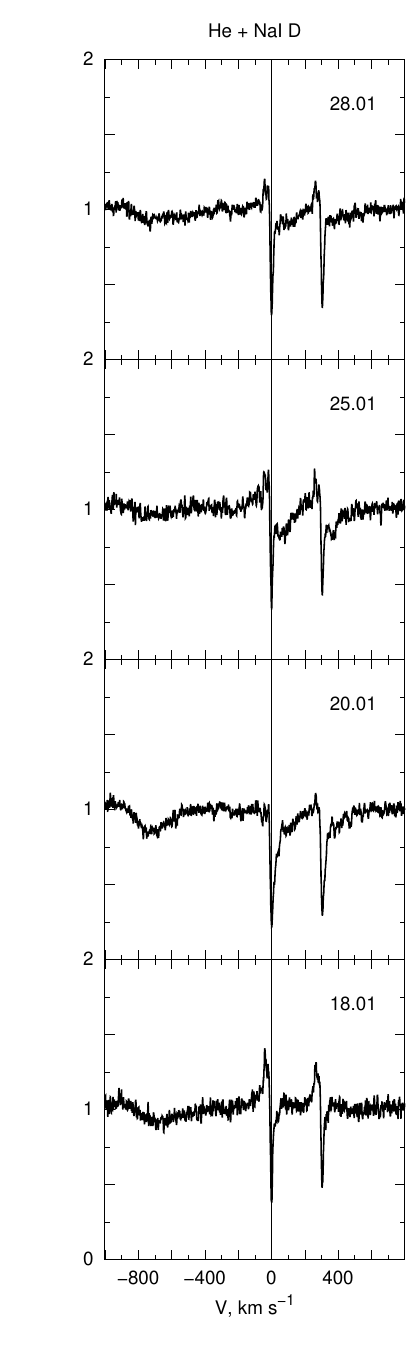}
\hspace{-0.6cm}\includegraphics[width=36mm]{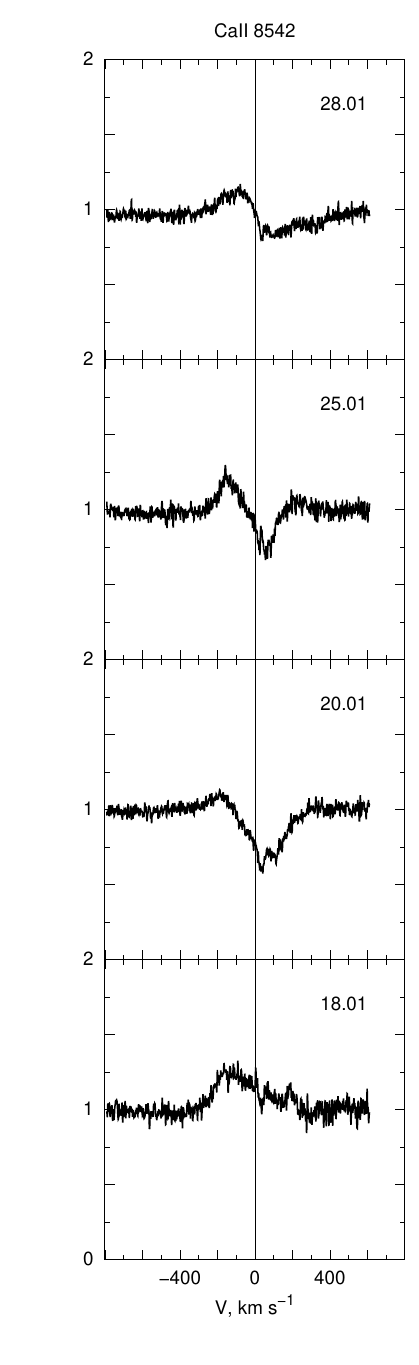}
\hspace{-0.6cm}\includegraphics[width=36mm]{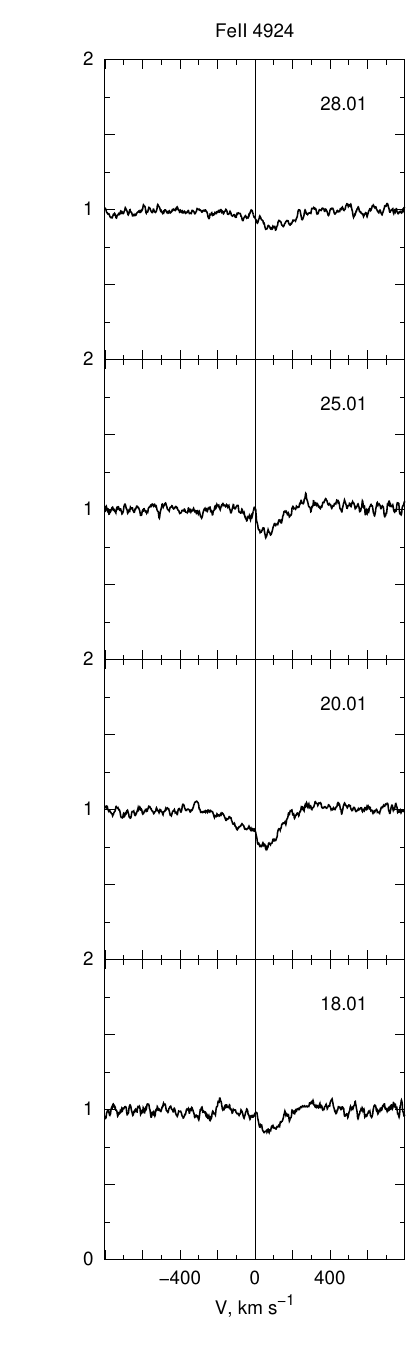}
\caption{The same as in Fig.~\ref{fig2}, but during brightness minimum in
  January 2021.}
\label{fig4}
\end{figure*}

\begin{figure*}
\hspace{-0.4cm}\includegraphics[width=36mm]{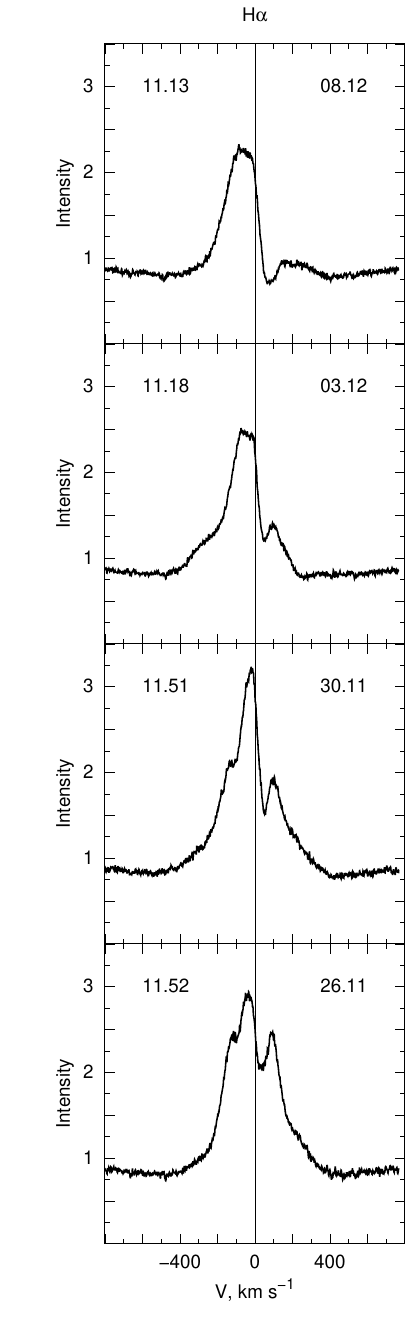}
\hspace{-0.6cm}\includegraphics[width=36mm]{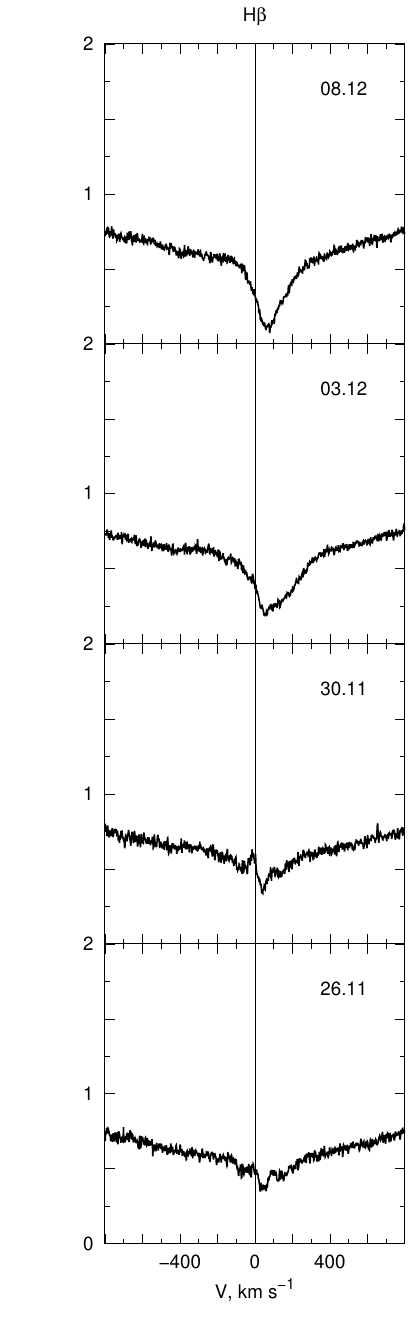}
\hspace{-0.6cm}\includegraphics[width=36mm]{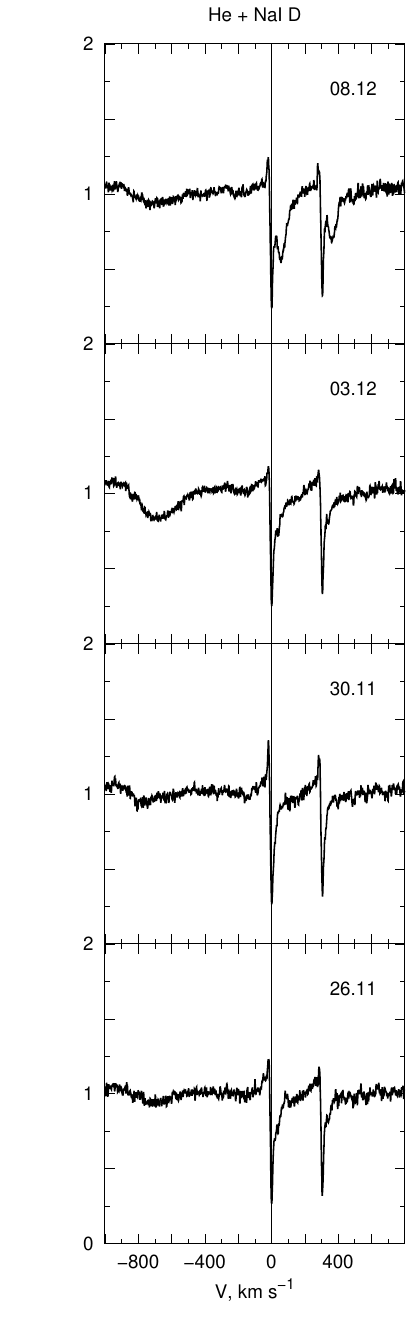}
\hspace{-0.6cm}\includegraphics[width=36mm]{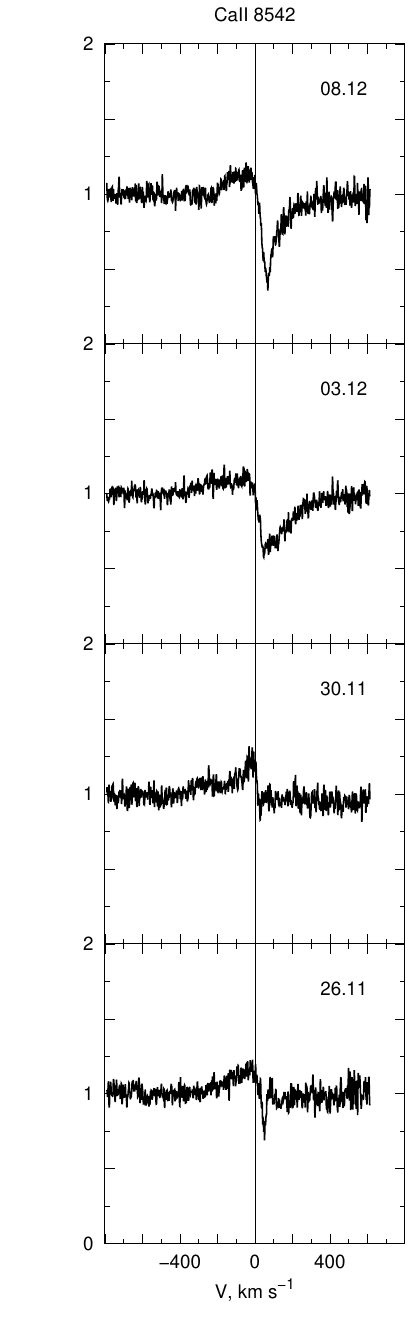}
\hspace{-0.6cm}\includegraphics[width=36mm]{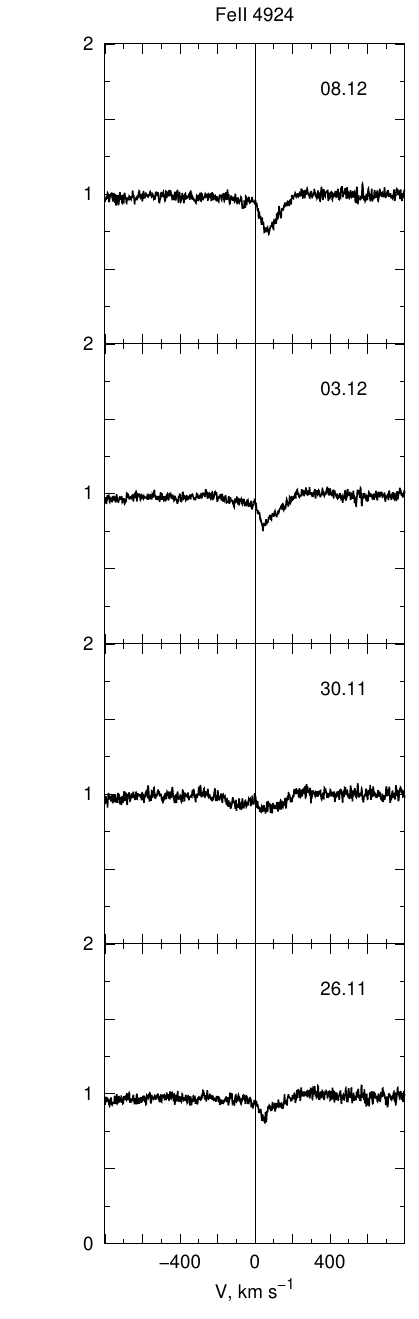}
\caption{The same as in Fig.~\ref{fig2} at the beginning of the brightness
  minimum in November-December 2023.}
\label{fig5}
\end{figure*}

\begin{figure*}
\hspace{-0.4cm}\includegraphics[width=36mm]{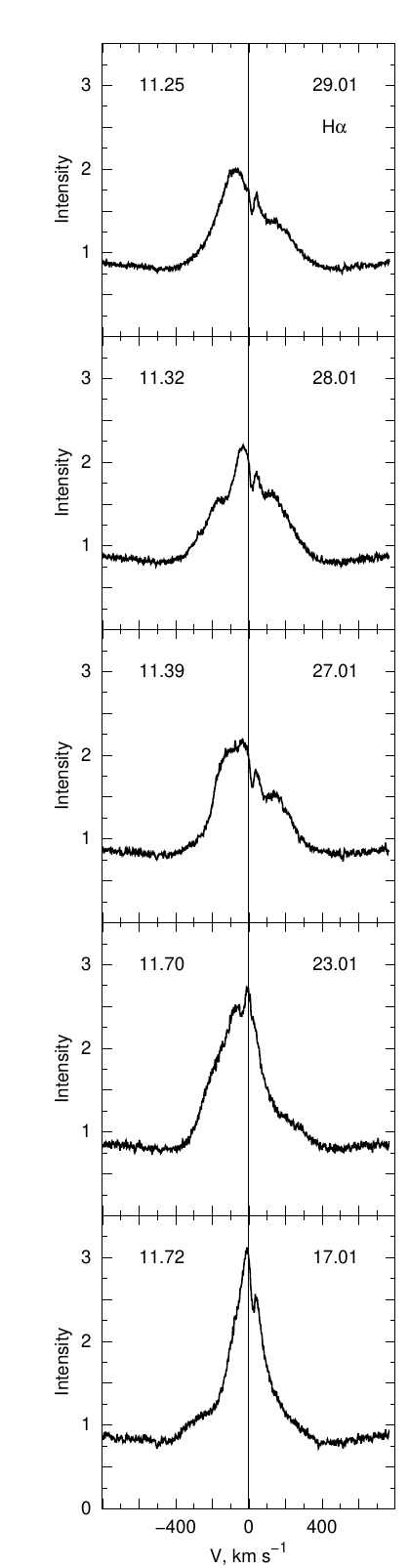}
\hspace{-0.6cm}\includegraphics[width=36mm]{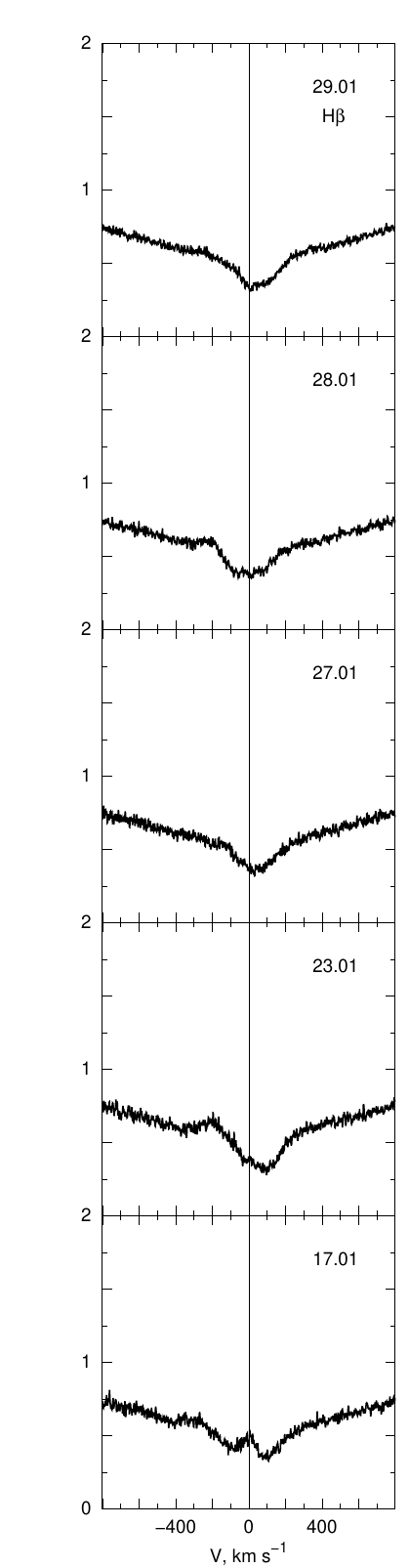}
\hspace{-0.6cm}\includegraphics[width=36mm]{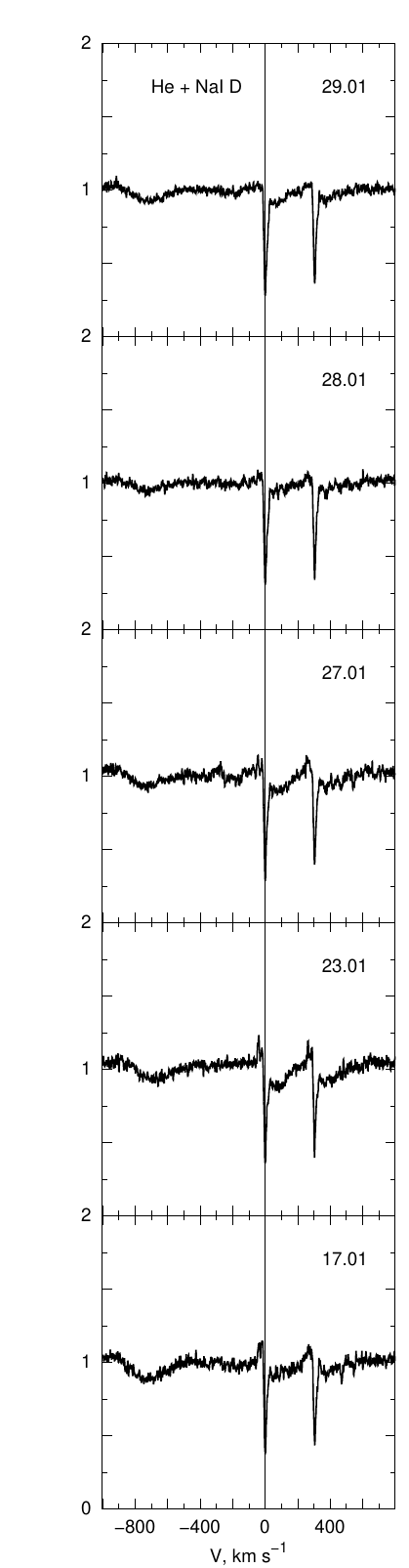}
\hspace{-0.6cm}\includegraphics[width=36mm]{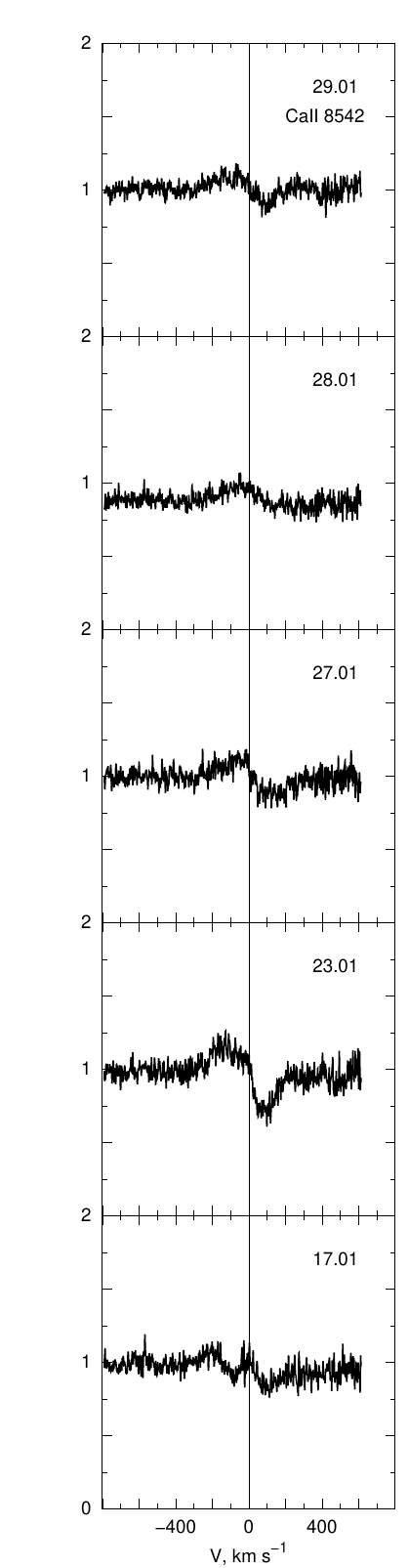}
\hspace{-0.6cm}\includegraphics[width=36mm]{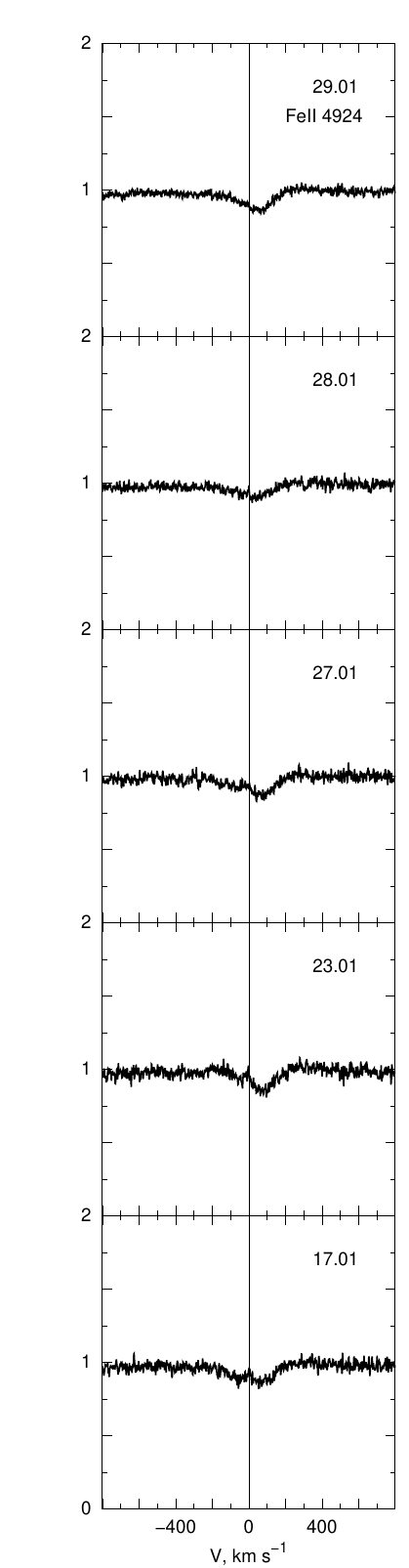}
\caption{The same as in Fig.~\ref{fig2}, but during the brightness
  minimum in January 2024.}
\label{fig6}
\end{figure*}

\begin{figure*}
\begin{centering}
\includegraphics[width=13cm]{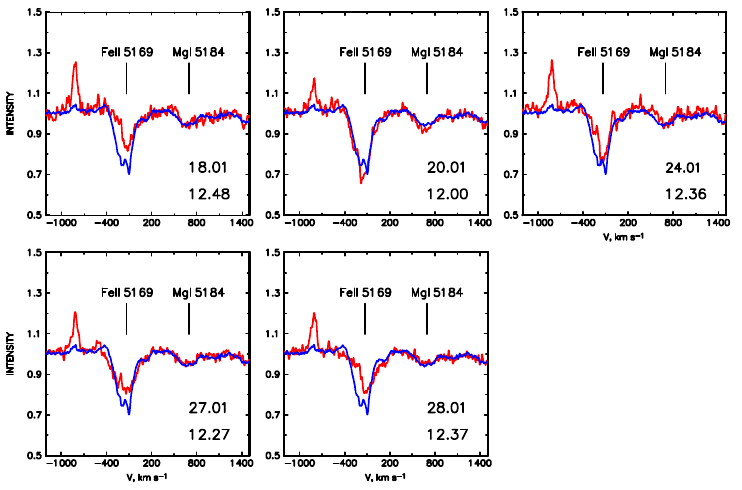}
\caption{MgI 5184 \AA \ and Fe II 5169 \AA \ in the spectra of
UX~Ori
  during the brightness minima in 2021 (red). The dates of the
  observations and the V-band magnitudes are shown in the each plot.
  The blue line represents the line profiles obtained in the bright state of
  the star on Sep. 5, 2020, with V= 9.98 mag.}
\label{fig20}
\end{centering}
\end{figure*}

\FloatBarrier

  \begin{table*}[t]
  \centering
  \caption{Spectra of UX~Ori and quasi-simultaneous V-band photometry.}
\newcommand\cola {\null}
\newcommand\colb {&}
\newcommand\colc {&}
\newcommand\cold {&}
\newcommand\cole {&}
\newcommand\colf {&}
\newcommand\colg {&}
\newcommand\eol{\\}
\newcommand\extline{&&&&&&\eol}

\begin{tabular}{lccccrl}
  \hline
  \hline
\cola Date (UT) FIES spectrum\colb $\chi$ \colc S/N\cold V'\cole $\sigma$\colf $\Delta t$\colg Photometry observer\eol
\cola \colb \colc \cold mag\cole mag\colf d\colg \eol
\hline
\cola 2019-02-20T23:04:15.0\colb 1.53\colc 49\cold 10.59\cole 0.10\colf -0.09\colg ASAS-SN (g) \eol
\cola 2019-02-23T21:47:21.4\colb 1.28\colc 78\cold 10.51\cole 0.03\colf -0.03\colg SAAF (V)\eol
\cola 2019-03-01T20:25:02.2\colb 1.20\colc 66\cold 10.81\cole 0.03\colf  0.05\colg SAAF (V)\eol
\cola 2019-03-03T21:18:07.2\colb 1.29\colc 37\cold 10.86\cole 0.10\colf -0.10\colg ASAS-SN (g)\eol
\cola 2019-03-05T21:34:48.5\colb 1.36\colc 52\cold 10.82\cole 0.10\colf -0.09\colg ASAS-SN (g)\eol
\cola 2019-03-06T22:01:13.8\colb 1.49\colc 58\cold 10.74\cole 0.10\colf  0.16\colg ASAS-SN (g)\eol
\cola 2020-01-30T01:27:28.0\colb 1.99\colc 34\cold 10.83\cole 0.10\colf -0.21\colg ASAS-SN (g)\eol
\cola 2020-02-15T20:43:10.6\colb 1.19\colc 25\cold 11.90\cole 0.10\colf -0.19\colg AAVSO (Visual)\eol
\cola 2020-03-09T20:52:41.6\colb 1.29\colc 38\cold 11.35\cole 0.10\colf  0.29\colg ASAS-SN (g)\eol
\cola 2020-03-12T23:05:02.6\colb 2.39\colc 22\cold 10.98\cole 0.03\colf -0.20\colg SAAF (V)\eol
\cola 2020-09-05T05:55:17.8\colb 1.30\colc 88\cold  9.98\cole 0.10\colf -0.72\colg ASAS-SN (g)\eol
\cola 2020-11-17T03:01:43.1\colb 1.19\colc 58\cold 10.89\cole 0.04\colf  0.01\colg SAAF (V)\eol
\cola 2020-12-07T00:18:23.1\colb 1.24\colc 43\cold 10.91\cole 0.01\colf  0.19\colg AAVSO (V)\eol
\cola 2020-12-12T03:12:57.1\colb 1.40\colc 26\cold 12.00\cole 0.10\colf -0.01\colg ASAS-SN (g)\eol
\cola 2020-12-18T00:23:02.5\colb 1.19\colc 46\cold 11.56\cole 0.01\colf  0.16\colg AAVSO (V)\eol
\cola 2020-12-21T01:05:00.0\colb 1.20\colc 59\cold 10.91\cole 0.02\colf  0.06\colg SAAF (V)\eol
\cola 2021-01-18T23:41:36.0\colb 1.23\colc 23\cold 12.48\cole 0.01\colf -0.07\colg SAAF (V)\eol
\cola 2021-01-20T21:43:29.3\colb 1.21\colc 34\cold 12.00\cole 0.10\colf  0.36\colg AAVSO (Visual)\eol
\cola 2021-01-24T21:23:19.5\colb 1.22\colc 26\cold 12.36\cole 0.04\colf -0.75\colg SAAF (V)\eol
\cola 2021-01-27T22:54:05.5\colb 1.22\colc 37\cold 12.27\cole 0.10\colf  0.23\colg ASAS-SN (g)\eol
\cola 2021-01-28T22:42:17.8\colb 1.21\colc 32\cold 12.37\cole 0.10\colf  0.23\colg ASAS-SN (g)\eol
\cola 2021-09-08T05:49:33.8\colb 1.29\colc 118\cold 10.30\cole 0.06\colf  0.16\colg SAAF (V)\eol
\cola 2021-09-20T04:53:59.2\colb 1.31\colc 86\cold 11.29\cole 0.09\colf  0.17\colg SAAF (V)\eol
\cola 2021-12-22T02:53:20.3\colb 1.48\colc 69\cold 11.28\cole 0.10\colf  1.14\colg ASAS-SN (g)\eol
\cola 2021-12-24T02:59:34.2\colb 1.55\colc 77\cold 10.61\cole 0.01\colf  0.66\colg SAAF (V)\eol
\cola 2021-12-25T00:44:22.0\colb 1.19\colc 88\cold 10.61\cole 0.02\colf -0.13\colg AAVSO (V)\eol
\cola 2021-12-28T22:52:44.9\colb 1.24\colc 90\cold 10.65\cole 0.10\colf  0.31\colg ASAS-SN (g)\eol
\cola 2021-12-30T03:35:07.5\colb 2.09\colc 62\cold 10.32\cole 0.07\colf  0.17\colg SAAF (V)\eol
\cola 2022-01-07T00:47:22.2\colb 1.30\colc 18\tablefootmark{a}\cold 10.30\cole 0.05\colf  0.13\colg SAAF (V)\eol
\cola 2022-01-15T21:37:03.2\colb 1.25\colc 83\cold 10.40\cole 0.01\colf -0.07\colg AAVSO (V)\eol
\cola 2022-02-14T20:42:20.8\colb 1.19\colc 69\cold 10.61\cole 0.05\colf  0.23\colg SAAF (V)\eol
\cola 2022-09-12T05:23:25.2\colb 1.32\colc 52\cold 10.5\cole  0.10\colf  0.0\colg V estimate from [OI] 6300 \AA \ line.\eol
\cola 2023-01-23T21:45:43.8\colb 1.20\colc 50\cold 10.60\cole 0.01\colf  0.18\colg AAVSO (V)\eol
\cola 2023-10-05T03:47:16.4\colb 1.33\colc 60\cold 10.43\cole 0.01\colf -0.01\colg NOT/ALFOSC (V)\eol
\cola 2023-10-17T02:22:45.6\colb 1.47\colc 55\cold 11.10\cole 0.01\colf  0.01\colg NOT/ALFOSC (V)\eol
\cola 2023-10-29T05:38:12.7\colb 1.32\colc 47\tablefootmark{a}\cold 10.86\cole 0.01\colf  0.04\colg NOT/StanCam (V)\eol
\cola 2023-11-02T03:09:46.9\colb 1.20\colc 61\cold 11.30\cole 0.01\colf -0.01\colg NOT/ALFOSC (V-)\eol
\cola 2023-11-26T01:23:21.0\colb 1.21\colc 45\cold 11.52\cole 0.01\colf  0.12\colg SAAF (V)\eol
\cola 2023-11-30T03:32:27.7\colb 1.31\colc 46\cold 11.51\cole 0.01\colf -0.01\colg NOT/ALFOSC (V)\eol
\cola 2023-12-03T02:22:34.9\colb 1.21\colc 59\cold 11.18\cole 0.01\colf  0.01\colg NOT/ALFOSC (V)\eol
\cola 2023-12-09T00:08:29.6\colb 1.24\colc 51\cold 11.13\cole 0.01\colf  0.01\colg NOT/ALFOSC (V)\eol
\cola 2024-01-11T22:32:32.7\colb 1.20\colc 33\cold 11.23\cole 0.01\colf -0.01\colg NOT/ALFOSC (V)\eol
\cola 2024-01-14T21:09:22.6\colb 1.32\colc 20\cold 11.72\cole 0.01\colf  0.01\colg NOT/ALFOSC (V)\eol
\cola 2024-01-17T20:53:54.1\colb 1.33\colc 39\cold 11.72\cole 0.01\colf  0.01\colg NOT/ALFOSC (V)\eol
\cola 2024-01-19T01:21:19.0\colb 1.60\colc 27\cold 11.67\cole 0.01\colf  0.02\colg NOT/ALFOSC (V)\eol
\cola 2024-01-19T22:13:58.6\colb 1.19\colc 22\cold 11.84\cole 0.01\colf  0.01\colg NOT/ALFOSC (V)\eol
\cola 2024-01-22T01:13:52.1\colb 1.64\colc 24\tablefootmark{a}\cold 11.75\cole 0.01\colf  0.07\colg AAVSO (V)\eol
\cola 2024-01-22T22:34:03.4\colb 1.19\colc 46\cold 11.77\cole 0.01\colf -0.06\colg NOT/StanCam (V)\eol
\cola 2024-01-23T22:54:14.9\colb 1.20\colc 41\cold 11.70\cole 0.01\colf -0.04\colg NOT/StanCam (V)\eol
\cola 2024-01-27T22:31:53.8\colb 1.20\colc 39\cold 11.39\cole 0.01\colf -0.01\colg NOT/ALFOSC (V)\eol
\cola 2024-01-28T22:21:18.3\colb 1.19\colc 51\cold 11.32\cole 0.01\colf  0.01\colg NOT/ALFOSC (V)\eol
\cola 2024-01-29T21:45:24.0\colb 1.19\colc 62\cold 11.25\cole 0.01\colf -0.01\colg NOT/ALFOSC (V)\eol
\cola 2024-02-03T00:22:07.5\colb 1.60\colc 10\cold 11.38\cole 0.01\colf  0.01\colg NOT/ALFOSC (V)\eol
\cola 2024-02-11T22:22:46.9\colb 1.27\colc 44\cold 11.87\cole 0.01\colf  0.01\colg NOT/ALFOSC (V)\eol
\hline
\end{tabular}
\tablefoot{The date refers to FIES spectrum, obtained at an airmass $\chi$, and with a final S/N ratio
  measured $\sim$ 5300 \AA. Quasi-simultaneous photometry is represented by the adopted V-band magnitude
  ($V'$), separated in time by $\Delta t$ fractional days from the FIES spectra. See Sect.~\ref{obs} for
  further details.   \\
  \tablefoottext{a}{FIES exposure time shorter than nominal.}
}
    \label{tab:fies-obs}
\end{table*}

\end{appendix}

\end{document}